\documentclass[a4paper,fleqn]{cas-sc}

\usepackage[numbers]{natbib}
\usepackage{microtype}
\usepackage{multirow}
\usepackage{tabularx}
\usepackage{tablefootnote}
\usepackage{hyperref}
\usepackage{minted}
\usepackage{amsmath}
\usepackage{float}
\usepackage{capt-of}

\def\tsc#1{\csdef{#1}{\textsc{\lowercase{#1}}\xspace}}
\tsc{WGM}
\tsc{QE}
\tsc{EP}
\tsc{PMS}
\tsc{BEC}
\tsc{DE}

\begin{document}
\let\WriteBookmarks\relax
\def\floatpagepagefraction{1}
\def\textpagefraction{.001}
\shorttitle{Balancing Privacy and Compliance}
\shortauthors{Li et~al.}

\title [mode = title]{Balancing Privacy and Compliance in DeFi: A Zero-Knowledge-Based Auditable Cross-Chain Framework}                      



\author[1]{Li Huiheng}[style=chinese,
                        orcid=0009-0004-3644-0451]
\credit{Conceptualization, Methodology, Project administration, Software, Writing - original draft}
\cormark[1]
\ead{Huiheng.Li@warwick.ac.uk}

\affiliation[1]{organization={Department of Computer Science, University of Warwick},
                city={Coventry},
                postcode={CV4 7AL}, 
                country={UK}}

\author[2]{Feng Kainuo}[
  style=chinese, 
  orcid=0009-0005-7591-2463]
\credit{Conceptualization, Formal analysis, Investigation, Methodology, Writing - Original draft}
\ead{fkn0530@student.gdufe.edu.cn}

\author[2]{Ding Jiahao}[%
   style=chinese, 
   orcid=0009-0005-1822-7662
   ]
\credit{Conceptualization, Formal analysis, Investigation, Methodology, Writing - Original draft}
\ead{19252060028@student.gdufe.edu.cn}

\affiliation[2]{organization={School of Public Finance and Taxation, Guangdong University of Finance and Economics},
                city={Guangzhou},
                country={China}}

\author[3]{Ma Ziqi}[%
   style=chinese, 
   orcid=0009-0002-8715-5576
   ]
\credit{Validation, Writing - review \& editing}

\ead{zctymad@ucl.ac.uk}

\affiliation[3]{organization={Department of Philosophy, University College London},
                city={London},
                postcode={WC1E 6BT}, 
                country={UK}}

\cortext[cor1]{Corresponding author}

\begin{abstract}
The rapid emergence of decentralized finance (DeFi) has introduced complex challenges for cross-chain transactions, which involve transferring assets across disparate blockchain networks. A fundamental dilemma arises between user privacy and ensuring regulatory compliance. Unlike single-chain systems, cross-chain environments must address privacy and auditability across heterogeneous architectures. Existing approaches,ranging from transparent ledgers to anonymous cryptocurrencies, fail to reconcile these two requirements, hindering regulatory adoption. This research presents an auditable cross-chain framework that integrates three core components. First, zero-knowledge proofs (ZKPs) enable compliance verification (e.g., amount non-negativity, signature validity) without revealing transaction details. Second, a light-client mechanism enables trust-minimized cross-chain verification without dependence on third-party relayers. Third, a threshold view-key mechanism based on distributed key generation (DKG) restricts that audit access to authorized entities under legal triggers such as the FATF Travel Rule and MiCA Regulation. For cross-border investigations, the framework adheres to national laws and the EU Directive on Mutual Legal Assistance. This work systematically integrates ZKPs, threshold cryptography, and light-client verification into an auditable, privacy-preserving cross-chain protocol, and evaluates a prototype on the Ethereum Sepolia testnet. The results demonstrate the practical feasibility of privacy-preserving compliance verification for cross-chain DeFi. At the same time, broader interoperability and Regulatory Technology (RegTech) impact require further validation on additional chains and with real regulatory workflows.
\end{abstract}



\maketitle

\section{Introduction}

Decentralized Finance (DeFi) and cross-chain transactions have grown exponentially over the past five years. The collective market capitalization of cryptocurrencies peaked at \$2 trillion in April 2021 \cite{lee_2021_crypto}. This expansion has fundamentally transformed financial infrastructure, enabling permissionless trading, lending, and asset transfers across blockchain networks. However, this growth has simultaneously created an acute conflict between two seemingly incompatible objectives: protecting user privacy and satisfying increasingly tight regulatory requirements \cite{deng_2025_a}. Traditional financial systems address this tension through centralized intermediaries that enforce know-your-customer (KYC) protocols and maintain audit trails. In contrast, decentralized systems must comply with regulations without sacrificing the privacy guarantees users expect from blockchain technology. \par
Current blockchain solutions represent two extremes \cite{feng_2025_secure}. Bitcoin and similar public ledgers prioritize transparency, recording all transactions on an immutable ledger accessible to any observer, a design that facilitates regulatory auditing but exposes sensitive financial information. Conversely, privacy-centric protocols such as Monero and Zcash apply advanced cryptographic techniques to shield transaction details, achieving strong privacy guarantees but rendering regulatory oversight nearly impossible. Neither approach provides a satisfactory middle ground: transparent systems leak private information, while privacy-preserving systems cannot selectively disclose transaction data when legally required. This incompatibility has hindered mainstream regulatory adoption of DeFi and cross-chain technologies, as financial authorities refuse to permit systems operating beyond their visibility. \par
Recent academic work has explored applying zero-knowledge (ZK) proofs to cross-chain protocols and privacy-preserving finance. However, existing research predominantly focuses on technical feasibility and performance optimization, overlooking a critical practical requirement: auditability under legal mandate. The literature lacks systematic frameworks that enable regulators to access encrypted transaction data only when presented with proper legal authorization (such as court orders or regulatory subpoenas), while maintaining strong privacy protections during normal operation \cite{chen_2024_crosschain}. This gap between academic theory and regulatory reality prevents the deployment of privacy-preserving DeFi systems in jurisdictions with active financial supervision. \par
To bridge this gap, we propose a framework that combines three building blocks. First, we adopt light-client-based cross-chain verification to achieve trust-minimized transaction validation. The target chain verifies source chain transactions via Merkle proofs without trusting any intermediary. Second, we integrate ZKPs to ensure transaction compliance (e.g., amount non-negativity, signature validity) without revealing transaction details. Third, we introduce a threshold view-key mechanism realized through a Pedersen distributed key generation (DKG) protocol over Shamir shares, which splits the audit decryption capability among $n$ regulators so that any $t$ of them can collaboratively decrypt the audit information, with decryption exercised only under the legal trigger conditions of \autoref{tab:trigger_scen}. This design ensures that during normal operation, transaction privacy is fully preserved; under lawful authorization, regulators can access only the minimum necessary audit information.\par
This work does not propose new cryptographic primitives; its contribution is a regulatory-oriented integration architecture in which privacy-preserving compliance verification (zero-knowledge proofs), trust-minimized cross-chain settlement (light clients), and threshold-controlled selective auditability cooperate within a single transaction lifecycle. Our principal contributions are:
\begin{enumerate}
  \item \textbf{Unified Regulatory-Compliant Cross-Chain Architecture.} We present an end-to-end system design that integrates zero-knowledge proofs (for privacy-preserving compliance verification), light-client verification (for trust-minimized cross-chain message passing), and threshold cryptography (for controlled audit access) into a single coherent framework. Unlike prior work that treats these components in isolation, our architecture jointly addresses privacy, auditability, and trust minimization as first-class design goals of a single orchestrated workflow.
  \item \textbf{Practical Auditable Cross-Chain Workflow.} We design and specify the full transaction lifecycle, from on-chain ZK proof verification and encrypted audit tag generation on the source chain, through light-client-based cross-chain asset release, to threshold-secured regulatory decryption under legal trigger conditions (FATF Travel Rule, MiCA Regulation, mutual legal assistance treaties). The encrypted audit label scheme stores only the minimum information regulators require, ensuring authorized access reveals precisely what is legally mandated and nothing more.
  \item \textbf{Prototype Implementation and Experimental Evaluation.} We implement the core components of the framework (Groth16 ZK circuits in Circom/snarkjs, light-client Solidity contracts on the Ethereum Sepolia testnet, and a Python-based threshold decryption simulator) and evaluate the prototype with synthetic transactions. The cross-chain verification component (audit-tag storage and light-client verification) adds approximately 170 ms of latency and 70k Gas, of the same order as a plain token transfer, on top of 214-229k Gas for on-chain Groth16 verification; these figures indicate feasibility for compliance-sensitive DeFi applications.
\end{enumerate}
The remainder of this paper is organized as follows: Section 2 reviews the related work. Section 3 presents the preliminaries and formalizes the problem. Section 4 details the system architecture and protocol design. Section 5 evaluates security properties. Section 6 assesses the experimental feasibility and performance. Section 7 discusses the current limitations. Section 8 concludes the research and proposes future research directions.

\section{Related Works}
Prior research relevant to our work spans three active areas: cross-chain communication (interoperability) protocols, privacy-preserving mechanisms for blockchain transactions, and regulatory compliance technologies. We review each in turn and identify where existing solutions partially address, but do not fully reconcile the tension between privacy and auditability. For cross-chain interoperability, we organize prior designs along a trust spectrum determined by whom a receiving chain must trust for an event that occurred on a sending chain: third-party relays/oracles and validator committees, on-chain light clients, and, finally, zero-knowledge proof bridges; for each rung we pinpoint the gap with respect to both privacy and auditability.

\subsection{Cross-Chain Interoperability: A Trust Spectrum}
\label{sec:crosschain}
Interoperability designs can be organized along a trust spectrum determined by whom a receiver chain must trust for a fact that occurred on a sender chain \cite{li_2024_blockchain,xie_2022_zkbridge}. At one end sit relay- and oracle-based architectures and validator committees, which entrust off-chain parties (or their signing keys) with observing source-chain events and attesting them on destination chains. The Axelar network, for instance, secures generalized message passing through threshold ECDSA signatures and validator voting \cite{nikolaeva_2022_axelar}. Security at this end is conditional on the honesty and custody of a small set of key holders: Wormhole relies on two-thirds of its guardian committee being honest \cite{xie_2022_zkbridge}, and Ronin lost five of its nine validator keys in a \$624M exploit \cite{xie_2022_zkbridge,liu_2024_accounting}; overall, bridge attacks extracted more than \$2.6B between 2021 and 2023 \cite{liu_2024_accounting}. An empirical accounting analysis of these incidents shows that the twelve largest attacks, although rooted in diverse implementation flaws, all manifest as unbalanced flows, withdrawals exceeding deposits less fees, and that a single balance invariant suffices to identify each one retrospectively and in real time \cite{liu_2024_accounting}. Financial-conservation auditing can therefore act as a safety layer orthogonal to whichever trust mechanism a bridge adopts; however, invariant-based monitoring presupposes fully transparent ledgers and offers neither privacy nor regulator-facing disclosure. In enterprise deployments of partially consistent ledgers (e.g., Corda and Fabric), a related design entrusts legally identified consortium gateways with contractual obligations, using signed "blockchain views" as stakeholder-centric, notarizable commitments of state for audits and cross-chain analysis \cite{belchior_2024_bungee}; accountability is achieved through multi-signature view endorsement, provenance metadata, and public notarization with dispute protocols, rather than through trust minimization.

In the middle of the spectrum, light-client protocols minimize trust by verifying cross-chain events directly against the source chain's consensus, typically through Merkle proofs validated against synchronized block headers; the Inter-Blockchain Communication (IBC) protocol is the canonical example of this paradigm \cite{zamyatin_2021_sok}. Such protocols remove third parties entirely, but they inherit the privacy properties (or lack thereof) of the underlying chains, offer no built-in selective disclosure, and become expensive or brittle when consensus is heterogeneous: verifying a single Proof-of-Stake header from Cosmos on Ethereum costs on the order of 64M gas \cite{xie_2022_zkbridge}, and light clients must additionally handle forks, long-range attacks, and validator rotation.

At the far end, zero-knowledge bridges compress the light-client verifier into a succinct argument, so a receiver chain accepts a constant-size proof of a source-chain state transition. In zkBridge, a permissionless block-header relay network the trusted only for liveness which submits headers together with zk-SNARKs to an updater contract that maintains the light-client state; security reduces to the soundness of the proof system, the security of the source chain, and the existence of a light-client protocol on it, with at least one honest relayer required only for liveness. Proof generation takes about 20 seconds for Cosmos headers using a distributed prover, proofs compress to 131 bytes via recursive verification, and on-chain verification costs under 230K gas \cite{xie_2022_zkbridge}. A ZK bridge certifies that a stated fact is authentic, not that value is conserved: accounting-level auditing remains necessary on top of even a trustless bridge \cite{liu_2024_accounting}, and no regulator-facing disclosure mechanism or accountable identity exists at this end of the spectrum.

Privacy-first intent-centric architectures such as Anoma move counterparty discovery to an open market of solvers while shielding settlement with zero-knowledge techniques; their open problem is private solving, i.e., letting solvers match intents without learning them (e.g., via trusted execution environments, multi-party computation, or collaborative SNARKs) \cite{khalniyazova_2023_anoma}. Privacy-preserving extensions to IBC ecosystems follow a similar philosophy of bolting privacy onto transparent transport. These lines of work treat auditability as an obstacle rather than a requirement and provide no channel for legally authorized selective disclosure.

Taken together, existing designs advance trust-minimized verification (IBC, zkBridge), transaction privacy (ZKP-based and intent-centric schemes), or auditability and accountability (chain analytics, invariant monitors, consortium views) largely in isolation. None simultaneously provides all three properties within a single cross-chain transaction; we revisit this three-dimensional gap in \autoref{sec:rwgap}, where \autoref{tab:rw_positioning} positions representative designs against our framework.

\subsection{Privacy-Preserving Blockchain Transactions}
Zero-knowledge proofs (ZKPs) have been applied to blockchain privacy across two complementary directions. On the scalability side, zkRollups aggregate off-chain transaction batches and submit a single succinct validity proof to the base layer, achieving throughput improvements while preserving some degree of input privacy \cite{sasson_2014_zerocash}. On the confidentiality side, protocols such as Zerocash employ zk-SNARKs to shield transaction amounts, sender, and recipient addresses, providing strong anonymity guarantees. These techniques have also been extended to cross-chain settings, where ZKPs can verify that a source-chain event occurred without revealing transaction details.

Anonymity-focused cryptocurrencies exhibit the same tension in a sharper and still-intensifying form. A systematic review of Monero's privacy architecture shows that upgrades such as FCMP++ extend the anonymity set to all historical outputs, undermining the statistical weaknesses of ring signatures that regulators previously exploited for heuristic tracing; simultaneously, tightening enforcement shifts liquidity from centralized venues toward harder-to-supervise decentralized markets. The review frames this as a \emph{privacy equilibrium}, in which regulatory pressure becomes an internal design variable of the system rather than an external constraint \cite{J2026Monero}. This dynamic indicates that auditability cannot be retrofitted onto a privacy-preserving system after the fact; it must be designed in from the start.

Notwithstanding these advances, existing privacy solutions face a common limitation when deployed in regulated environments. zkRollups protect individual transaction inputs but leave aggregate patterns (e.g., trading volumes, fee structures) amenable to statistical analysis; fully shielded protocols such as Zerocash provide cryptographic-grade anonymity that inherently conflicts with Anti-Money Laundering (AML) and Know-Your-Customer (KYC) obligations \cite{amlnetwork_2025_kyc}. In all cases, the protocol lacks a mechanism for selective, legally authorized disclosure: either too much information is visible, or none at all. This binary trade-off has limited the adoption of privacy techniques in compliance-sensitive cross-chain applications.

\subsection{Regulatory Compliance and Auditing}
On the regulatory side, the FATF's revised guidance for Virtual Asset Service Providers (VASPs) mandates that customer identity information accompany fund transfers, the so-called Travel Rule \cite{fatf_2021_updated}. Traditional blockchain analytics platforms, such as those developed by Chainalysis, address this need by constructing complete transaction graphs through address clustering and heuristic labeling to identify suspicious flows \cite{chainalysis_2024_blockchain}. These tools have proven effective for post-hoc forensic investigation and risk scoring.

Nevertheless, current compliance tools exhibit two structural tensions with privacy-preserving architectures. First, their effectiveness depends on access to unencrypted, complete transaction data, which is fundamentally incompatible with shielded or zero-knowledge-based transaction designs. Second, their centralized data collection model introduces a single point of information aggregation, creating risks of mass data leakage and unauthorized surveillance.

A recent wave of work has begun to reconcile privacy with auditability directly in cross-chain settings; these proposals can be organized by \emph{what is disclosed, to whom, and under what cryptographic control}. A first family enforces \textbf{threshold-authorized decryption}: transaction metadata is encrypted or secret-shared so that a committee can reveal it only under a $t$-of-$n$ consensus. Mystiko layers a $t$-out-of-$n$ privacy committee, whose members hold shares of the decryption key and never the whole key, over ZK-rollup confidential transfers, so that a quorum of auditors, and not any single party, can open a transaction \cite{Li2025An}. VeilAudit's identity-reveal phase similarly assembles threshold ElGamal shares across a multi-party authority before de-anonymizing an entity \cite{Qiao2025VeilAudit}. A second family provides \textbf{auditor-only linkability} without identity: VeilAudit's auditor-only linkability (AOL) combines Groth16 ownership proofs with public-key encryption supporting equality tests (PKE-ET), letting an auditor determine whether two pseudonymous tags originate from the same entity across chains while remaining unable to recover the underlying identity \cite{Qiao2025VeilAudit}. A third family yields \textbf{privacy-preserving aggregate audit proofs}: privXCA organizes a dedicated audit chain and a distributed proving network so that a regulator verifies a single recursive Plonky2 proof certifying batch-level, cross-chain, and time-window compliance, observing only aggregate amounts and counts rather than individual transactions \cite{Wang2026privXCA}. A fourth family confines accountability to \textbf{intermediaries rather than users}: notary-scheme designs equip a notary group with traceable group signatures, allowing a supervisory group to open and punish a misbehaving notary. In contrast, user transactions remain unlinkable, which means that auditing the users themselves is explicitly left as future work \cite{Wu2025Privacy}. Related relay-chain and cross-chain audit proposals also attach privacy-preserving audit receipts to relay or bridge architectures. Still, they either retain trust in the relaying layer \cite{feng_2025_secure}, stop short of a fully specified authorized-decryption protocol \cite{Jiang2025Cross}, perform integrity and correctness auditing of cross-chain data exchanges via bilinear pairings without zero-knowledge compliance checks or regulator-facing disclosure \cite{deng_2025_a}, or rely on a single trusted auditor who decrypts complete transaction details without any threshold control \cite{chen_2024_crosschain}. \autoref{tab:rw_audit} positions these designs against our framework.

\begin{table}[t]
  \caption{Positioning of recent privacy-preserving auditable cross-chain schemes against our framework. \checkmark{} = provided; $\triangle$ = partial; $\times$ = absent. \emph{Sel.\ audit} = regulator-facing selective disclosure; \emph{Trust-min.} = trust-minimized verification. The distinguishing axis is the audit paradigm and its target.}
  \label{tab:rw_audit}
  \small
  \begin{tabularx}{\textwidth}{|l|c|c|c|c|c|X|}
  \toprule
  \textbf{Design} & \textbf{Cross-chain} & \textbf{Privacy} & \textbf{ZKP} & \textbf{Sel.\ audit} & \textbf{Trust-min.} & \textbf{Audit paradigm / object} \\
  \midrule
  \midrule
  VeilAudit \cite{Qiao2025VeilAudit} & \checkmark & \checkmark & \checkmark & $\triangle$ & $\triangle$ & Auditor-only linkability; threshold identity reveal (user) \\ \hline
  privXCA \cite{Wang2026privXCA} & \checkmark & \checkmark & \checkmark & $\triangle$ & $\triangle$ & Aggregate compliance proofs (user, batch-level) \\ \hline
  Mystiko \cite{Li2025An} & \checkmark & \checkmark & \checkmark & \checkmark & $\triangle$ & Threshold-authorized decryption (user) \\ \hline
  CPAP \cite{Jiang2025Cross} & \checkmark & \checkmark & \checkmark & $\triangle$ & $\times$ & Authorized entity, underspecified (user) \\ \hline
  Notary + ZKP \cite{Wu2025Privacy} & \checkmark & \checkmark & \checkmark & $\triangle$ & $\triangle$ & Traceable group signatures (intermediary only) \\ \hline
  Deng et al. \cite{deng_2025_a} & \checkmark & $\triangle$ & $\times$ & $\times$ & $\triangle$ & Bilinear-pairing integrity audit (data exchange; not regulator-facing) \\ \hline
  Chen et al. \cite{chen_2024_crosschain} & \checkmark & \checkmark & \checkmark & $\triangle$ & $\triangle$ & Single trusted auditor decrypts full tx (no threshold) \\ \hline
  Our framework & \checkmark & \checkmark & \checkmark & \checkmark & \checkmark & Threshold decrypt + minimal info (user) \\
  \bottomrule
  \end{tabularx}
\end{table}

\subsection{Research Gap and Motivation}
\label{sec:rwgap}
\autoref{tab:rw_positioning} and \autoref{tab:rw_audit} position the designs reviewed above against our framework. In summary, existing work has achieved significant progress within each of the three areas: cross-chain communication (trust-minimized verification), privacy-preserving finance (ZKPs for transaction confidentiality), and regulatory technology (blockchain analytics and compliance frameworks). However, these solutions generally operate in isolation, specifically, they address at most one or two of the three properties simultaneously. To the best of our knowledge, no existing framework jointly provides: (i) trust-minimized cross-chain verification, (ii) cryptographic transaction privacy via zero-knowledge proofs, and (iii) controlled auditability via threshold-secured selective disclosure. This paper proposes an integrated architecture that systematically combines these three building blocks, addressing a concrete gap between the technical capabilities of decentralized finance and the operational requirements of regulated financial supervision.

\begin{table}[t]
  \caption{Positioning of cross-chain interoperability designs discussed in \autoref{sec:crosschain}. \checkmark{} = provided; $\triangle$ = partial; $\times$ = absent. \emph{Sel.\ audit} = regulator-facing selective disclosure; \emph{Trust-min.} = trust-minimized verification.}
  \label{tab:rw_positioning}
  \small
  \begin{tabular}{|l|c|c|c|c|c|}
  \toprule
  \textbf{Design} & \textbf{Cross-chain} & \textbf{Privacy} & \textbf{ZKP} & \textbf{Sel.\ audit} & \textbf{Trust-min.} \\
  \midrule
  \midrule
  Relay/committee designs (Axelar) \cite{nikolaeva_2022_axelar} & \checkmark & $\times$ & $\times$ & $\times$ & $\times$ \\ \hline
  IBC (light-client paradigm) \cite{zamyatin_2021_sok} & \checkmark & $\times$ & $\times$ & $\times$ & \checkmark \\ \hline
  zkBridge \cite{xie_2022_zkbridge} & \checkmark & $\times$ & \checkmark & $\times$ & \checkmark \\ \hline
  BUNGEE \cite{belchior_2024_bungee} & \checkmark & $\triangle$ & $\times$ & $\triangle$ & $\triangle$ \\ \hline
  Invariant auditing \cite{liu_2024_accounting} & $\triangle$ & $\times$ & $\times$ & $\triangle$ & $\triangle$ \\ \hline
  Anoma \cite{khalniyazova_2023_anoma} & $\triangle$ & \checkmark & $\triangle$ & $\times$ & $\triangle$ \\ \hline
  Our framework & \checkmark & \checkmark & \checkmark & \checkmark & \checkmark \\
  \bottomrule
  \end{tabular}
\end{table}

\section{Preliminaries and Problem Formalization}

\subsection{Cryptographic Preliminaries}
\subsubsection{Zero-Knowledge Proofs}
\label{section:zkproperty}
Zero-Knowledge Proofs (ZKP) is a type of cryptographic protocol that allows prover to demonstrate the truth of a statement to a verifier without revealing any additional information. Formally, a ZKP scheme consists of three algorithm groups:
\begin{itemize}
  \item $\text{Setup} (1^\lambda) \rightarrow pp$: Generate public parameters.
  \item $\text{Prove} (pp, x, w) \rightarrow \pi$: Input a public statement $x$ and a private witness $w$, and generate proof $\pi$.
  \item $\text{Verify} (pp, x, \pi) \rightarrow \{0,1\}$: Verify if $\pi$ is valid.
\end{itemize}
ZKP has the following characteristics:
\begin{itemize}
  \item \textbf{Completeness}: If $w$ is correct, the validation passes.
  \item \textbf{Soundness}: If $w$ is incorrect, the probability that verification accepts is negligible.
  \item \textbf{Zero-knowledge}: The validator cannot obtain any information about $w$ from $\pi$.
\end{itemize}
This research adopts the Groth16 scheme. Users use ZKP to prove transaction compliance (such as a positive amount and a valid signature) without exposing specific transaction details.

\subsubsection{Threshold Secret Sharing}
Threshold secret sharing schemes allow a secret $s$ to be split into $n$ fragments $s_1,...,s_n$, such that any $t$ fragments can be used to recover $s$, while any $t-1$ fragments cannot be used to obtain any information.\par
For share generation stage, given a secret $s$, threshold $t$, and number of participants $n$, the dealer randomly generates a polynomial $f(x)=s+\sum ^{t-1} _{j=1}a_j x^j$ of degree $t-1$. Each participant $i$ receives a fragment $s_i=f(i)$. For reconstruction, given any $t$ fragments ${(i,s_i)}^t _{i=1}$, the secret is recovered via Lagrange interpolation:
\begin{equation}
  s = \sum^t _{i=1} s_i \cdot \prod_{j \neq i}\frac{-j}{i-j} \text{.}
\end{equation} 
This scheme ensures information-theoretic security: fewer than $t$ fragments provide zero information about $s$ . Our framework employs Shamir's Secret Sharing as the underlying primitive for threshold key management; the shares are produced in a decentralized manner by the Pedersen DKG protocol of \autoref{sec:dkg}, so that the full secret never exists at any single point.

\subsubsection{Light Client and State Proofs}
Lightweight clients verify the existence of transactions using Merkle proofs without downloading the complete blockchain. Specifically: a) Each block header contains either a ${state}_{root}$ or a ${tx}_{root}$ (the Merkle root of the transaction tree). b) A transaction $tx$ and its position $pos$ can be proven to belong to a block via the path $path$: 
\begin{equation}
  \text{MerkleVerify} (\text{tx}_\text{root}, \text{tx}, \text{path}, \text{pos}) \rightarrow \{0,1\} \text{.}
\end{equation}

\subsection{System Model}
The system consists of the following entities:
\begin{itemize}
  \item \textbf{User}: The individual or institution initiating the cross-chain transaction. The user holds a source chain account $addr_s$ and a target chain account $addr_t$, and generates a ZK proof for the transaction.
  \item \textbf{Source Chain $C_s$}: The blockchain on which the user initiates the transaction. An audit contract $AuditContract$ is deployed on it, responsible for verifying the ZK proof and storing the cryptographic audit tag.
  \item \textbf{Target Chain $C_t$}: The blockchain receiving the asset. A light client contract $LightClient$ (continuously verifies the source chain block header) and an execution contract $ExecContract$ (responsible for releasing the asset) are deployed on it.
  \item \textbf{Regulatory Set $\mathcal{R} = \{R_1, \ldots, R_n\}$}: A group of regulatory bodies that jointly hold the threshold view key. Any $t$ regulators can collaborate to decrypt audit information. 
  \item \textbf{Relayer (Optional)}: Submits the source chain's block header and Merkle proof to the target chain. In a trustless design, anyone can play this role by participating through Gas incentives.
\end{itemize}

\subsection{Threat Model and Assumptions}
\subsubsection{Trust Assumptions}
\label{sec:assumptions}
In this work, we require specific trust assumptions: a) the consensus mechanism of the underlying blockchain is secure, meaning that confirmed blocks cannot be tampered with. b) Cryptographic primitives (including hash functions, elliptic curves, and ZKP) are computationally secure, without considering extreme cases of potential breaches or undiscovered vulnerabilities. c) The DKG protocol execution involves an honest majority of regulators (i.e., fewer than $t$ malicious participants), which is the standard security assumption for threshold cryptography. d) Regulatory nodes strictly adhere to threshold decryption protocols and related decryption procedures, the decryption communications are neither monitored nor tampered with, and that decrypted keys exist only on temporary storage devices and are promptly destroyed after the compliance process is completed. e) Deployment scope: the current prototype targets EVM-compatible chains; extending the framework to non-EVM chains (e.g., Bitcoin, Solana) requires chain-specific light-client and Merkle-proof implementations, which is out of scope for the present evaluation.\par
Under the current assumptions, this work aims to prevent the following threats: a) users attempting to submit illegitimate transactions or bypass audit tag generation. b) External attackers attempting to infer transaction details from publicly available on-chain data (ZK proofs, cryptographic tags, etc.). c) A single regulator attempting to decrypt audit information without authorization, or a total of no more than $t$ regulators colluding. \par
This work does not prevent more than $t$ regulators maliciously colluding to decrypt audit information without authorization, the breach of the underlying chain consensus, or the leakage of user private keys or supervisory private keys.

\subsubsection{Adversary Model}
\label{sec:adversary}
We formalize the capabilities and limitations of a probabilistic polynomial-time (PPT) adversary $\mathcal{A}$ against our framework:

\noindent\textbf{Capabilities.} The adversary $\mathcal{A}$ can:
\begin{enumerate}
  \item \textbf{Observe} all on-chain data, including ZK proofs $\pi$, encrypted audit tags $C$, block headers, and smart contract state on both the source and target chains.
  \item \textbf{Corrupt} up to $t-1$ regulators, obtain their key fragments $sk_{i}$, and observe all their internal states and communications.
  \item \textbf{Interact} with the protocol as a user, submitting arbitrary transactions to the source chain and relay requests to the target chain.
  \item \textbf{Act} as a network-level adversary, eavesdropping, delaying, or dropping messages exchanged between honest regulators.
\end{enumerate}

\noindent\textbf{Limitations.} The adversary $\mathcal{A}$ cannot:
\begin{enumerate}
  \item Break the \textbf{computational soundness} of the ZK proof system (Groth16).
  \item Break the \textbf{semantic security} of the ECIES encryption scheme (i.e., cannot solve DDH).
  \item Break the \textbf{collision resistance} of the underlying hash function.
  \item Compromise the \textbf{consensus security} of the underlying blockchain (i.e., cannot forge or revert confirmed blocks).
  \item Corrupt $t$ or more regulators simultaneously.
\end{enumerate}

This adversary model is standard for threshold-cryptographic systems and captures a powerful but computationally bounded attacker who may control a minority of regulators but cannot subvert the core cryptographic assumptions.

\subsubsection{Security Goals}
\label{sec:secgoals}
Under the adversary model defined in \autoref{sec:adversary}, our protocol $\Pi$ aims to satisfy the following three security goals:

\noindent\textbf{Goal 1 (Transaction Privacy).} For any two transactions $\text{tx}_0, \text{tx}_1$ that satisfy the same compliance predicate $P$, the adversary $\mathcal{A}$ cannot distinguish which transaction was executed based on the protocol's public output. Formally:
\begin{equation}
  |\Pr[\mathcal{A} (\pi (\text{tx}_0), C_0) = 1] - \Pr[\mathcal{A} (\pi (\text{tx}_1), C_1) = 1]| \leq \mathsf{negl}(\lambda) \text{,}
\end{equation}
where $\pi(\text{tx})$ is the ZK proof and $C = \mathsf{Enc}_{pk_R}(M_{min}(\text{tx}))$ is the encrypted audit tag.

\noindent\textbf{Goal 2 (Controlled Auditability).} This goal comprises a cryptographic guarantee and an authorization guarantee. \emph{Cryptographic guarantee}: fewer than $t$ key-share holders cannot decrypt the audit information regardless of legal status, whereas any $t$ of them can recover $M_{min}(\text{tx})$ by computing valid partial decryption shares $d_i = \mathsf{DecShare}(sk_i, C)$. \emph{Authorization guarantee}: the exercise of this decryption capability is gated by the legal trigger conditions of \autoref{tab:trigger_scen} (e.g., judicial warrant, regulatory investigation), which are enforced through off-chain institutional procedures governing the release and combination of decryption shares and are recorded in the on-chain audit log. In particular, any coalition of fewer than $t$ regulators (even if all are corrupted by $\mathcal{A}$) obtains negligible information about $M_{min}(\text{tx})$.

\noindent\textbf{Goal 3 (Collusion Resistance).} A coalition of up to $t-1$ corrupted regulators cannot recover the audit plaintext $M_{min}(\text{tx})$, even if they pool all their key fragments $sk_i$ and all publicly available on-chain data. This follows directly from the information-theoretic security of Shamir's Secret Sharing.

\subsubsection{Threat Model Analysis}
To systematically identify the security boundaries and potential attack vectors of our architecture, we employ the STRIDE threat modeling framework. \autoref{tab:stride} maps each threat category to specific cross-chain audit scenarios and our corresponding defense mechanisms.
\begin{table*}[!htbp]
  \centering
  \caption{STRIDE Threat Analysis of the Proposed Framework}
  \label{tab:stride}
  \begin{tabularx}{\textwidth}{|l|X|X|}
  \toprule
\textbf{STRIDE Category} &
  \textbf{Threat Scenario} &
  \textbf{Defense Mechanism} \\
  \midrule
  \midrule
Spoofing (S) &
  Adversaries forge cross-chain transactions or impersonate legitimate users/relayers to submit invalid proofs. &
  Digital signatures verified within the ZK circuit ensure transaction origin authenticity. Light client contracts verify Merkle proofs to guarantee the validity of cross-chain messages. \\ \hline
Tampering (T) &
  Man-in-the-middle attacks altering transaction details or adversaries modifying encrypted audit labels on-chain. &
  ZKP binding ensures proof integrity. The collision resistance of cryptographic hashes and encryption schemes protects audit labels from undetected modification. \\ \hline
Repudiation (R) &
  A user denies initiating a cross-chain transaction, or a regulator denies participating in a decryption process. &
  Non-interactive ZKPs with digital signatures provide user non-repudiation; the (t, n)-threshold decryption protocol requires t regulators to provide verifiable key fragments, ensuring operational non-repudiation. \\ \hline
Information Disclosure (I) &
  External adversaries inferring transaction details from on-chain data, or fewer than t regulators colluding to decrypt audit information. &
  ZKP's zero-knowledge property hides transaction details from the public; the $(t,n)$-threshold scheme combined with ECIES encryption ensures that fewer than $t$ shares reveal negligible information. \\ \hline
Denial of Service (D) &
  Targeted DDoS attacks on the network infrastructure of regulatory nodes, or prolonged offline status of regulators, preventing the assembly of t decryption shares. &
  Audit unavailability is an accepted architectural trade-off: the $(t, n)$-threshold scheme intentionally prevents unilateral decryption, meaning that if $\ge n-t+1$ regulators are incapacitated, audit liveness is sacrificed to preserve privacy. \\ \hline
Elevation of Privilege (E) &
  A standard user attempting to gain regulatory decryption privileges to access audit information. &
  Inherently prevented. The mathematical structure of (t, n)-threshold cryptography ensures that users without a valid key share cannot escalate privileges through computation.\\
  \bottomrule
\end{tabularx}
\end{table*}
As shown in \autoref{tab:stride}, our framework inherently mitigates Spoofing, Tampering, Repudiation, and Elevation of Privilege through cryptographic assumptions, while specifically addressing Information Disclosure and balancing Denial of Service risks via the (t, n)-threshold scheme. To rigorously specify the security guarantees against the identified threats (particularly I and R), we formalize the adversary model and security goals in \autoref{sec:adversary}-\ref{sec:secgoals} and provide formal security definitions in \autoref{section:problem_form}.

\subsubsection{Distributed Key Generation Security}
Key generation is decentralized by design: the framework replaces any centralized key-generation center (KGC) with the Pedersen DKG protocol of \autoref{sec:dkg}, so the full private key $sk_R$ never exists at any single location and there is no dealer whose compromise would expose it. An adversary that corrupts a regulator's device during initialization obtains only that regulator's share $sk_i$ and the broadcast commitments, which are insufficient to decrypt any audit tag without $t$ shares. For completeness, we also bound the damage of a hypothetical KGC-based variant: under a key-erasure assumption (the KGC deletes all key material after distribution), a post-distribution compromise yields no useful information, and a compromise \textit{during} initialization yields only the encrypted audit tags $C = \mathsf{Enc}_{pk_R} (M)$, which remain computationally secure under the DDH assumption. In either case, decrypting historical transactions still requires $\ge t$ regulator shares.

\subsection{Problem Formalisation}
\label{section:problem_form}
\subsubsection{Definition of Cross-Chain Transaction}
A cross-chain transaction can be represented as a tuple:
\begin{equation*}
\begin{split}
  \text{tx} = (\text{txHash}_\text{s}, \text{timestamp}_\text{i}, \text{timestamp}_\text{c}, \text{asset}, \text{amount},  \text{VASP}_\text{s}, \text{addr}_\text{s}, \text{internalID}_\text{s},\\ \text{VASP}_\text{t}, \text{addr}_\text{t}, \text{internalID}_\text{t}, \text{status}, \text{STRid}, \text{VASP}_\text{transit}, \text{addr}_\text{transit} ) \text{,}
\end{split}
\end{equation*}
in \autoref{tab:audit_info}.
\begin{table*}
  \caption{Minimum Auditing Necessary Information}
  \label{tab:audit_info}
  \begin{tabularx}{\textwidth}{|c|X|c|X|}
    \toprule
    \textbf{Info Type} &
    \textbf{Specific fields} &
    \textbf{Element Name} &
    \textbf{Legal basis} \\
    \midrule
    \midrule
    Transaction core identifier\footnotemark[1] &
    Transaction hash &
    $\text{txHash}_\text{s}$ &
    FATF R.16 / MiCA  \cite{fatf_2025_fatf} \\
    \midrule
  \multirow{2}{*}{Time information\footnotemark[2]} &
    Transaction initiation timestamp &
    $\text{timestamp}_\text{i}$ &
    6AMLD / AML laws in various countries \cite{ivxsuklimited_2020_6amld} \cite{deane_2025_a} \\ 
  &
    Transaction confirmation timestamp &
    $\text{timestamp}_\text{c}$ &
    \\
    \midrule
  \multirow{2}{*}{Asset Information\footnotemark[3]} &
    Asset types (e.g., BTC, ETH) &
    $\text{asset}$ &
    \multirow{2}{*}{FATF R.15-16 \cite{chainalysisteam_2024_fatfs} \cite{fatf_2025_fatf}} \\
  &
    Amount &
    $\text{amount}$ &
    \\
    \midrule
  \multirow{4}{*}{Trading direction\footnotemark[4]} &
    Source VASP Name &
    $\text{VASP}_\text{s}$ &
    FATF R.16 (2021 Revised Guidelines) \cite{fatf_2025_fatf} \\
  &
    Source address &
    $\text{addr}_\text{s}$ &
    \\
  &
    Terminal VASP Name &
    $\text{VASP}_\text{t}$ &
    \multirow{2}{*}{MiCA $\S$63 (5)} \\
  &
    Terminal address &
    $\text{addr}_\text{t}$ &
    \\
    \midrule
  \multirow{2}{*}{Account ID\footnotemark[5]} &
    Source VASP Internal Account ID &
    $\text{internalID}_\text{s}$ &
    AML Enforcement Rules in various countries \cite{amlnetwork_2025_aml} \\
  &
    Terminal VASP Internal Account ID &
    $\text{internalID}_\text{t}$ &
    \\
    \midrule
  \multirow{2}{*}{Compliance metadata} &
    Travel rule information delivery status\footnotemark[6] &
    $\text{status}$ &
    FATF R.16 technical compliance \cite{fatf_2025_fatf} \\
  &
    Suspicious Transaction Report number &
    $\text{STRid}$ &
    6AMLD $\S$25 \cite{ivxsuklimited_2020_6amld} \\
    \midrule
  \multirow{2}{*}{Technical path\footnotemark[7]} &
    Transit VASP name &
    $\text{VASP}_\text{transit}$ &
    FATF Virtual Asset Guidance \cite{fatf_2025_asset} \\
  &
    Transit address &
    $\text{addr}_\text{transit}$ &
  \\
  \bottomrule
\end{tabularx}
\end{table*}

\footnotetext[1]{Used for on-chain verification and tracking}
\footnotetext[2]{Judging the timeliness of transactions}
\footnotetext[3]{Assess risk level}
\footnotetext[4]{Determine the responsible party}
\footnotetext[5]{Anonymous identifiers are mapped to users only when required by law (and can be linked to KYC)}
\footnotetext[6]{Verify compliance collaboration between VASPs}
\footnotetext[7]{Used for tracing complex transaction chains}

The work focuses on \textbf{compliance-sensitive transactions}, which are the types of transactions that regulators need to audit under specific conditions.

\subsubsection{Definition of Security Properties}
Building on the adversary model $\mathcal{A}$ and security goals formalized in \autoref{sec:adversary}-\ref{sec:secgoals}, we now present the formal security definitions that our protocol $\Pi$ must satisfy.

\noindent\textbf{Definition 1 (Transaction Privacy).}
Let $\mathcal{A}$ be any PPT adversary as defined in \autoref{sec:adversary}. For any two transactions $\text{tx}_0, \text{tx}_1$ that satisfy the same compliance predicate $P$, the protocol's public output $(\pi, C)$ must be computationally indistinguishable:
\begin{equation}
  |\Pr[\mathcal{A} (\pi_0, C_0) = 1] - \Pr[\mathcal{A} (\pi_1, C_1) = 1]| \leq \mathsf{negl}(\lambda) \text{,}
\end{equation}
where $\pi_b = \Pi.\mathsf{Prove}(pp, x_b, w_b)$ is the ZK proof whose public input contains the transaction commitment $c_b = \mathsf{Commit}(\text{tx}_b)$, $C_b = \mathsf{Enc}_{pk_R}(M_{min}(\text{tx}_b))$ is the encrypted audit tag whose plaintext records $\text{txHash}_s := c_b$, and $\mathsf{negl}(\lambda)$ is a negligible function in the security parameter $\lambda$; the proof and the tag are thus bound to the same commitment $c_b$. This definition captures the requirement that the adversary learns nothing beyond the fact that a compliant transaction occurred.

\noindent\textbf{Definition 2 (Controlled Auditability).}
For any transaction $\text{tx}$, let $M_{min}(\text{tx})$ be its minimum audit information set. The protocol must satisfy:
\begin{enumerate}
  \item \textbf{Threshold Recovery (cryptographic layer)}: Any $t$ regulators, each providing a valid partial decryption share $d_i = \mathsf{DecShare}(sk_i, C)$, recover $M_{min}(\text{tx})$:
  \begin{equation}
    \mathsf{Combine}(d_1, \ldots, d_t) \rightarrow M_{min}(\text{tx}) \text{.}
  \end{equation}
  \item \textbf{Unauthorized Secrecy}: For any coalition $\mathcal{C} \subset \{R_1, \ldots, R_n\}$ with $|\mathcal{C}| < t$, even if all members of $\mathcal{C}$ are corrupted by $\mathcal{A}$, the advantage in recovering any information about $M_{min}(\text{tx})$ is negligible in the security parameter $\lambda$.
\end{enumerate}
The protocol does \emph{not} cryptographically verify whether the legal trigger conditions of \autoref{tab:trigger_scen} hold; compliance with legal authorization is enforced through the off-chain institutional procedures that govern when regulators may compute and combine decryption shares, and every decryption event is recorded in the on-chain audit log.

\noindent\textbf{Definition 3 (Trust-Minimized Cross-Chain Verification).}
The target chain $C_t$ verifies the inclusion of a source-chain transaction $\text{tx}$ in block $B$ using only:
\begin{enumerate}
  \item The block header $H_B$ of $B$ (validated against the consensus of $C_s$);
  \item A Merkle proof $\pi_M$ proving that $\text{tx}$ is included in $H_B.\mathsf{txRoot}$.
\end{enumerate}
The verification $\mathsf{MerkleVerify}(H_B.\mathsf{txRoot}, \text{tx}, \pi_M, \text{pos})$ relies on no trusted third party and no economic collateral; its correctness follows solely from the collision resistance of the underlying Merkle tree hash function.

\noindent\textbf{Definition 4 (Proof-Tag Binding).}
Let $c = \mathsf{Commit}(\text{tx})$ be a binding commitment to the canonical encoding of the transaction fields (instantiated by a collision-resistant hash of the normalized transaction tuple). A valid protocol execution satisfies:
\begin{enumerate}
  \item the public input of the ZK statement includes $c$, i.e., $x = (c, \ldots)$, and the circuit enforces that the private witness $w$ matches $c$;
  \item the minimum audit information encrypted in the audit tag records $\text{txHash}_s := c$, i.e., $C = \mathsf{Enc}_{pk_R}(M_{min}(\text{tx}))$ with $c \in M_{min}(\text{tx})$;
  \item the audit tag metadata sets $\mathsf{Tag.txHash} := c$ and $\mathsf{Tag.zkProofHash} := H(\pi)$, so that the stored proof, the commitment, and the ciphertext are publicly checkable as belonging to the same transaction.
\end{enumerate}
This binding ensures that decrypting a tag yields the audit information of exactly the transaction whose compliance was verified, preventing a regulator from being misled by a proof and a tag generated from different transactions.

\subsubsection{Compliance Predicates and Audit Information Definitions}
The compliance predicate $P (\text{tx})$ represents the legal conditions that a transaction must meet, including but not limited to a non-negative transaction amount, a valid transaction signature, and a transaction amount not exceeding the anti-money laundering threshold (which triggers requiring regulatory audit).

The \textit{Minimum Auditing Necessary Information} set $M_{min} (\text{tx})$ is defined as:
\begin{equation}
  M_{min} (\text{tx}) = \{ \text{txHash}_\text{s}, \text{timestamp}_\text{i}, \text{timestamp}_\text{c}, \text{asset}, \text{amount}, ... \}  \text{.} \label{eq:m_min}
\end{equation}
See \autoref{tab:audit_info} for specific fields. This set represents the minimum necessary information required for regulatory audits and complies with requirements such as the FATF Travel Rules.

\subsubsection{Core Problem Statement}
In summary, the core problem addressed in this paper is to design a cross-chain transaction protocol such that:
\begin{enumerate}
  \item Compliance can be publicly verified via ZK proofs, while transaction details remain private.
  \item The \textit{Minimum Auditing Necessary Information} of a transaction is recorded on-chain in encrypted form, decryptable only by a set of regulatory entities that meet legal triggering conditions.
  \item The cross-chain verification process is fully trustless, and implemented using light-client state proofs.
\end{enumerate}

\section{System Architecture and Protocol Design}
\subsection{Overall Architecture}
Our work proposes a privacy-protected and auditable cross-chain transaction architecture.

\begin{center}
    \includegraphics[width=.9\textwidth]{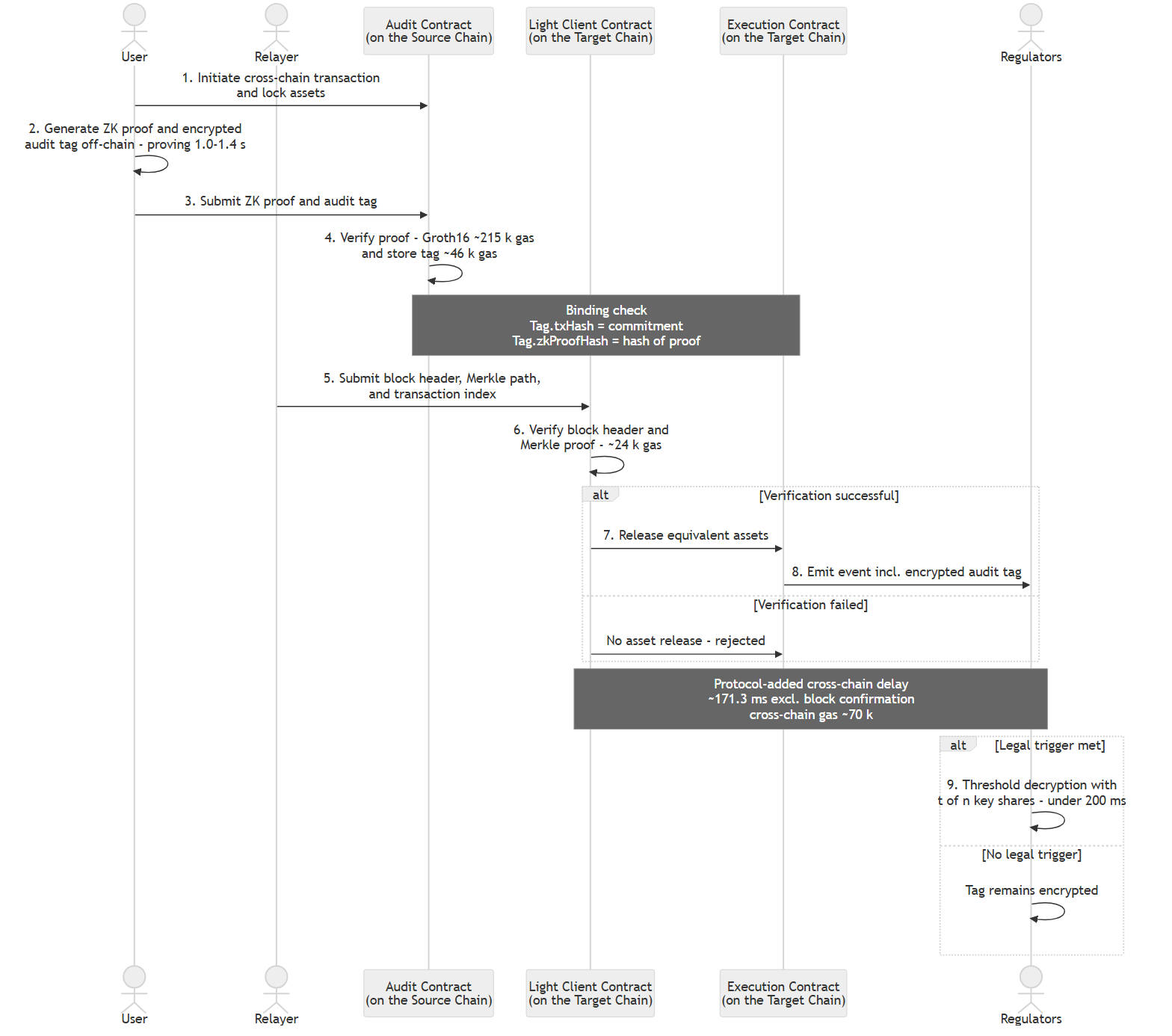}
    \captionof{figure}{Transaction and Audit Workflow}
    \label{fig:sys_archi}
\end{center}

As \autoref{fig:sys_archi} shows, the system consists of three layers:
\begin{itemize}
  \item \textbf{User Layer}: The transaction initiator, responsible for constructing the transaction, generating zero-knowledge proofs, and cryptographic audit tags.
  \item \textbf{On-Chain Contract Layer}: Includes the audit contract deployed on the source chain (verifying proofs and storing audit tags), the light client contract (verifying cross-chain messages), and the execution contract (releasing assets) deployed on the target chain.
  \item \textbf{Regulatory Layer}: Composed of $n$ regulatory bodies, holding the critical $ (t,n)$-threshold decryption private key, collaborating only to decrypt audit information when legal triggering conditions are met.
\end{itemize}
Its core workflow is as follows:
\begin{enumerate}
  \item A user initiates a transaction on the source chain, locks assets, and generates a ZK proof and a cryptographic audit tag off-chain.
  \item The user submits the proof and tag to the source chain audit contract. The contract verifies the validity of the proof and stores the tag.
  \item The target chain's light client verifies the existence of the transaction on the source chain using a Merkle proof.
  \item Upon successful verification, the target chain executes the contract to release assets of equivalent value.
  \item When legal triggering conditions are met, regulators collaborate to decrypt the audit tag and obtain compliance audit information.
\end{enumerate}

\subsection{Transaction Generation Process (Source Chain)}
ZK circuit framework employs zero-knowledge proofs to ensure transaction compliance without exposing transaction details. We design a ZK circuit $\mathcal{C}$, whose inputs include: \textbf{Private Inputs} (Known by the prover, not publicly disclosed): Sender's private key signature, transaction amount, recipient's address, and the normalized transaction fields. \textbf{Public Inputs} (Stored on-chain): the transaction commitment $c = \mathsf{Commit}(\text{tx})$, the blacklist root $root_{BL}$, and the compliance predicate output value.
The circuit verifies the following assertion:
\begin{itemize}
  \item \textbf{Signature Validity}: The transaction is signed by a legitimate sender.
  \item \textbf{Amount Validity}: The transaction amount is non-negative and does not exceed the account balance.
  \item \textbf{Compliance Mark}: When the transaction amount exceeds a threshold, it is marked as \textbf{Audit Required}.
  \item \textbf{Commitment Binding}: The witness matches the public commitment $c$: the circuit recomputes $\mathsf{Commit}(\text{tx})$ from the private transaction fields and asserts equality with the public input $c$.
\end{itemize}
If all assertions are true, the circuit outputs $\pi = \text{Prove} (pp, x, w)$, where $x$ is the public input and $w$ is the private witness. We instantiate the proof with Groth16, chosen for its constant-size proofs ($\approx$800 bytes) and its constant on-chain verification cost, both of which suit resource-constrained smart contracts; the measurement of \autoref{sec:zkperf} resolves comparisons against PLONK/STARK: benchmarking the snarkjs PLONK prover on the same circuits yields per-transaction proving $\approx$13-17$\times$ slower and serialized proofs $\approx$2.8$\times$ larger, confirming Groth16 as the practical choice for per-transaction compliance proofs; optimized PLONK implementations (e.g., Plonky2) remain future work. Three methods reduce constraint overhead. (i) All in-circuit hashing, the transaction commitment and the blacklist Merkle tree, uses the ZK-friendly Poseidon permutation rather than SHA-256 or MiMC, sharply reducing the nonlinear constraints per hashed input. (ii) Range and threshold checks are implemented with circomlib's comparison-based \emph{Num2Bits}/\emph{LessThan} components instead of arithmetic encodings. (iii) The blacklist is verified through a sorted-Poseidon Merkle non-membership proof (Section 4.4.3), so its cost grows logarithmically rather than linearly in the blacklist size.\par

To support controlled auditing, each transaction must generate an encrypted audit tag $\mathsf{Tag}$. Its plaintext contains the minimum audit information set $M$ as defined in Equation (\ref{eq:m_min}), with $\text{txHash}_s$ set to the transaction commitment $c$, thereby binding the tag to the same transaction that the ZK proof certifies. The specific fields comply with FATF travel rules. The user encrypts $M$ using the regulatory public key $pk_R$:
\begin{equation}
  C = \mathsf{Enc}_{pk_R} (M) \text{.}
\end{equation}
The encryption algorithm uses the Elliptic Curve Integrated Encryption Scheme (ECIES), outputting ciphertext $C$. The final audit tag structure is as follows:
\begin{minted}[linenos, frame=lines, framesep=2mm]{Solidity}
struct AuditTag {
    bytes ciphertext;
    bytes32 txHash;
    bytes32 zkProofHash;
    uint8 threshold;
    bytes32[] regulatorIds;
}
\end{minted}
The metadata fields make the binding publicly checkable: $\mathsf{txHash}$ stores the commitment $c$ (equal to the commitment in the public input of the verified proof), and $\mathsf{zkProofHash}$ stores $H(\pi)$, so that any observer can match the stored tag to the exact proof verified on-chain.

The user submits the ZK proof $\pi$ and the audit tag $\mathsf{Tag}$ to the source chain audit contract $\mathsf{AuditContract}$. Contract execution:
\begin{enumerate}
  \item Verify $\mathsf{Verify} (pp, x, \pi)$ with $x = (c, \ldots)$, rolling back if it fails.
  \item Check the binding: $\mathsf{Tag.txHash} = c$ and $\mathsf{Tag.zkProofHash} = H(\pi)$; roll back otherwise.
  \item If all checks pass, store $\mathsf{Tag}$ in the on-chain state.
  \item Trigger the event $\mathsf{AuditTagStored} (\mathsf{Tag.txHash})$ for the target chain to listen for.
\end{enumerate}
This design ensures that only compliant transactions are recorded on the chain, that audit information is permanently stored in encrypted form, and that each stored tag is provably bound to the very proof that certified the transaction's compliance.

\subsection{Cross-Chain Verification Phase (Light Client)}

To maintain block header synchronization, the target chain deploys a light client contract, $\mathsf{LightClient}$, from the source chain to continuously synchronize the source chain's block headers. The light client verifies the validity of the PoW/PoS consensus for each block header and maintains a verifiable block header. This mechanism does not require trust in any third-party relay.\par
Once a user's transaction $\text{tx}$ on the source chain is confirmed, anyone (the relayer) can submit a verification request to the target chain, including the block header $H$ of the transaction, the Merkle path $\text{path}$ of the transaction in the block, and the transaction index $\text{pos}$. Light client contract verification: a) Whether $H$ is already in the synchronized block header chain. b) Whether $\mathsf{MerkleVerify} (\text{H.txRoot}, \text{tx}, \text{path}, \text{pos})$ is true. \par
If the verification passes, it proves that the $\text{tx}$ indeed exists on the source chain. After successful verification, the light client contract invokes the execution contract $\mathsf{ExecContract}$, releasing equivalent assets to the user's address on the target chain. Simultaneously, the audit tag $\mathsf{Tag}$ is copied to the target chain's storage, ensuring that audit information remains attached regardless of which chain the transaction flows to.

\noindent\textbf{Data Availability and Relay Incentives.}
Light-client verification introduces a data availability requirement: block headers and Merkle proofs must be explicitly submitted by a relayer for the target chain to verify a cross-chain transaction. This differs from relay-chain architectures (e.g., Feng et al. \cite{feng_2025_secure}), which maintain continuous data feeds but require trust in the relay network. Our framework mitigates the data availability issue through an economic incentive mechanism: any user or third party can serve as a relayer and receive gas compensation from the execution contract upon successful proof submission. As long as the cross-chain transaction value exceeds the gas cost of proof submission, rational relayers are incentivized to act. This preserves the trust-minimization property of light-client verification without sacrificing liveness in practice.

\subsection{Controlled Audit Designs (Regulatory Layer)}
\subsubsection{Distributed Key Generation}
\label{sec:dkg}
The work assumes a regulatory set $\mathcal{R} = \{R_1, R_2, ..., R_n\}$, consisting of the following regulatory bodies: \textbf{Financial Intelligence Unit (FIU)}, \textbf{Financial Supervisory Authority (FSA)}, \textbf{Industry Self-Regulatory Organization (ISRO)}, \textbf{Foreign Regulatory Agencies} (through mutual legal assistance treaties/ memoranda) and \textbf{Judicial Department} (including Police and Prosecutors' Offices, involving criminal offenses). To eliminate the trusted initialization assumption, we replace the centralized KGC with a DKG protocol among the $n$ regulators. Specifically, we adopt the Pedersen DKG protocol, which allows the regulators to jointly generate a public key $pk_R$ and secret shares $sk_1,...,sk_n$ without any party ever learning the full private key $sk_R$. The DKG is executed once during system initialization; no further interaction among regulators is required for key management thereafter. The protocol proceeds as follows:
\begin{enumerate}
  \item Each regulator $R_i$ broadcasts a commitment to a random polynomial $f_i(x)$ of degree $t-1$.
  \item After all commitments are received, each $R_i$ sends a share $f_i(j)$ to every other $R_j$.
  \item Each $R_j$ verifies received shares against the broadcast commitments.
  \item The final public key is $pk_R=\prod_i g^{f_i(0)} $ , and regulator $R_j$'s secret share is $sk_j=\sum_i f_i(j)$.
\end{enumerate}
This DKG protocol ensures that no single entity ever possesses the full decryption key, eliminating the KGC as a single point of trust. The security of this protocol holds under the Discrete Logarithm assumption, with up to $t-1$ malicious regulators tolerated during generation. 
\subsubsection{Audit Trigger Scenarios}
According to regulatory policies and the legal framework, the declassification of audit information is only triggered in the scenarios listed in \autoref{tab:trigger_scen}.
\begin{table*}
  \caption{Triggering Scenarios}
  \label{tab:trigger_scen}
  \begin{tabularx}{\textwidth}{|X|X|X|X|}
    \toprule
  \textbf{Scenario} &
  \textbf{Main Responsible Bodies} &
  \textbf{Triggering Conditions} &
  \textbf{Information Scope} \\ \midrule \midrule
Regulatory proactive investigation &
  FIU &
  STR triggers &
  Minimal audit information (excluding user identity) \\ \hline
FSA Review &
  FSA &
  Regular compliance inspections or spot checks &
  Minimal audit information + VASP compliance record \\ \hline
VASP Internal Risk Alert &
  VASP Compliance Department &
  VASP proactively reports suspicious behavior. &
  Transaction metadata + VASP information \\ \hline
Cross-border collaboration request &
  Foreign regulatory agencies &
  Foreign regulatory agencies through mutual legal assistance &
  Minimum necessary information + legal basis statement \\ \hline
Law enforcement agencies require &
  Police/Prosecutor's Office &
  Criminal investigation (with a judicial warrant) &
  Minimal audit information + user identity information \\ \bottomrule
\end{tabularx}
\end{table*}
The relevant audit triggering and execution details are:\par
Based on STR or intelligence analysis and after internal approval, the FIU issues intelligence requests to the VASP. Prior to the operation, it is necessary to confirm that the scope of the request is directly related to legal duties, such as suspected money laundering and terrorist financing \cite{schwindwagner_2025_fatf}. After the FSA provides the VASP with compliance background and preliminary risk assessment, the audit process is witnessed by the ISRO to ensure the completeness of the data extraction and the due process.\par
The FSA may initiate an audit based on on-site inspections, off-site monitoring, or whistleblowing, subject to approval from its internal law enforcement department \cite{fatfoecd_2012_international} \cite{europeanparliament_2023_regulation}. Intelligence support is provided by the FIU to confirm the presence of suspicious activity, and the ISRO provides an independent opinion and cooperates with the investigation.\par
VASP's compliance department identifies suspicious transactions or high-risk behaviors through its transaction-monitoring system \cite{europeanparliament_2018_directive}. VASP's internal risk control system triggers an alert, automatically generating a risk report and submitting it through the regulatory-designated system \cite{theeuropeancommission_2023_commission}. If the suspicion is confirmed, a tripartite assessment of the alert level determines whether to initiate an audit. VASP's compliance department does not hold the security key. It can only submit risk reports. If the current situation triggers a suspicious investigation, an audit will be jointly initiated by the ISRO, FIU, and FSA.\par
If a cross-border joint investigation is involved, a formal request from the foreign regulatory agency is required, along with a statement of legal basis \cite{europeanparliament_2014_directive}. If a Mutual Legal Assistance Treaty (MLAT) exists, cross-border audit applications can be submitted through the Ministry of Foreign Affairs to the Supreme Court/Supreme Procuratorate and then to the corresponding foreign agency as criminal judicial assistance \cite{officeofinternationalaffairscriminaldivisionusdepartmentofjustice_2022_mutual}. In another way, the exchange of regulatory information is limited to non-criminal matters and handled by the domestic financial regulatory authority in accordance with the Multilateral Memorandum of Understanding (MMoU) when it receives foreign regulatory requests \cite{internationalorganizationofsecuritiescommissions_2016_2013}. This request must be translated into a formal document from a competent domestic authority (such as a financial regulatory body or court) before it can be obtained from the VASP \cite{officeofinternationalaffairscriminaldivisionusdepartmentofjustice_2022_mutual}. The legality and reciprocity of the request are reviewed by the domestic judicial authorities, and the FIU assesses the request's compatibility with domestic law \cite{thecommissionoftheeuropeancommunities_2000_2000800ec}.\par
When initiating criminal cases involving suspected money laundering, illegal business operations, or fraud, judicial personnel, armed with a \textit{Notice of Evidence Collection} and their work identification, request an audit \cite{homeoffice_2024_0052024}. When collecting evidence, judicial authorities need to verify fulfillment of VASP compliance obligations, thereby requiring the FSA to conduct its work \cite{financialservicesandthetreasurybureauofthehongkongspecialadministrativeregiongovernment_2018_antimoney}. The FIU's investigative leads also need to cooperate with the audit.

\subsubsection{Dynamic Blacklist Management with Merkle Tree}
A critical challenge is that compliance predicates (e.g., VASP blacklists) evolve. A naive ZK circuit that hardcodes the blacklist would invalidate all previous proofs upon each update. To address this, we treat the blacklist as a Merkle tree whose root $root_{BL}$ is maintained on-chain by a designated authority (e.g., FIU). The ZK circuit verifies membership or non-membership by checking a Merkle proof against the current $root_{BL}$. The circuit input includes:
\begin{itemize}
  \item Private input: The VASP identifier $id$ being checked.
  \item Public input: The current blacklist root $root_{BL}$ and a Merkle proof $\pi _{BL}$.
\end{itemize}
The circuit verifies $\text{MerkleVerify}(root_{BL},\pi _{BL})$ to determine if id is in the blacklist. Because $root_{BL}$ is a public input, it can change between transactions without affecting previously generated proofs, since each proof is bound to the $root_{BL}$ at the time of transaction submission. This design ensures:
\begin{itemize}
  \item Forward compatibility: New blacklist entries do not retroactively invalidate old proofs (the compliance check is against the blacklist state at transaction time, which is the correct semantic).
  \item Light client sync: The target chain's light client can synchronize the latest $root_{BL}$ by reading the source chain's state root, ensuring cross-chain consistency.
\end{itemize}

\subsubsection{Threshold Decryption and Key Fragment Management}
\noindent\textbf{Combining ECIES, Shamir sharing, and Pedersen DKG.} The three primitives are assembled as follows. During system initialization, the $n$ regulators run Pedersen DKG (\autoref{sec:dkg}), which jointly produces the regulatory public key $pk_R$ and hands each $R_i$ a Shamir share $sk_i$ of the implicit secret $sk_R$; no party ever learns $sk_R$ itself. At transaction time, the user encrypts the audit plaintext $M$ under $pk_R$ with ECIES, obtaining a ciphertext $C = (C_1, C_2)$ whose ephemeral public component is $C_1 = g^r$ and whose payload $C_2$ is the symmetric encryption of $M$ under a key derived from the shared Diffie-Hellman secret $pk_R^r$. Decryption is then performed in a threshold manner \emph{in the exponent}. When the triggering conditions are met, a corresponding number of regulators collaborate:
\begin{enumerate}
  \item Each regulator $R_i$ reads the encrypted audit tag $C$ from the blockchain and computes a \textbf{partial decryption share} $d_i = \mathsf{DecShare}(sk_i, C) = C_1^{sk_i}$ using its private key fragment $sk_i$.
  \item The partial decryption shares $d_i$ are exchanged via a secure channel (TLS + mutual authentication).
  \item Once at least $t$ valid partial decryption shares are collected, the shared secret $pk_R^r = \prod_{i \in S} d_i^{\ell_i}$ is recovered by combining the shares in the exponent with Lagrange coefficients $\ell_i$; the symmetric key is derived, and the plaintext $M$ is decrypted, i.e., $\mathsf{Combine}(d_1, \ldots, d_t) \rightarrow M$. The full private key $sk_R$ is \textbf{never reconstructed} during this process.
  \item \textbf{Binding verification}: the regulators check that the recovered plaintext satisfies $\text{txHash}_s = c = \mathsf{Tag.txHash}$, confirming that the decrypted audit information belongs to exactly the transaction whose compliance was certified on-chain.
\end{enumerate}
Each regulator's key fragments are stored in a Hardware Security Module (HSM), requiring dual authentication for access. During decryption, partial decryption shares must reside only on the runtime temporary storage device and must be securely deleted after the audit and forensics process concludes. No decryption-related material may persist in any device in any form. All decryption operations, including the release of shares and the binding verification result, are recorded in the on-chain audit log, achieving "audit of audits" and preventing abuse of power.

\section{Security Analysis}
In this section, we provide formal security arguments for our protocol $\Pi$ with respect to the adversary model $\mathcal{A}$ and security goals defined in \autoref{sec:adversary}-\ref{sec:secgoals}. We structure the analysis as a series of theorems, each accompanied by a proof sketch.

\subsection{Transaction Privacy}

\begin{description}
\item[Theorem 1 (Transaction Privacy).] Under the assumptions that (i) the Groth16 ZK proof system satisfies computational zero-knowledge, and (ii) the ECIES encryption scheme provides semantic security under the DDH assumption, for any PPT adversary $\mathcal{A}$ as defined in \autoref{sec:adversary} and any two transactions $\text{tx}_0, \text{tx}_1$ satisfying the same compliance predicate $P$:
\begin{equation}
  |\Pr[\mathcal{A}(\pi_0, C_0) = 1] - \Pr[\mathcal{A}(\pi_1, C_1) = 1]| \leq \mathsf{negl}(\lambda) \text{.}
\end{equation}
\end{description}

\noindent\textit{Proof Sketch.}
We proceed via a hybrid argument.
\begin{enumerate}
  \item \textbf{ZK Hybrid}: By the computational zero-knowledge property of Groth16, there exists a simulator $\mathcal{S}$ that produces simulated proofs $\pi^{\mathsf{sim}}$ computationally indistinguishable from real proofs. Replacing $\pi(\text{tx}_0)$ with $\pi^{\mathsf{sim}}$ changes $\mathcal{A}$'s advantage by at most $\mathsf{negl}_1(\lambda)$.
  \item \textbf{Encryption Hybrid}: By the semantic security of ECIES under DDH, $C_0 = \mathsf{Enc}_{pk_R}(M_0)$ and $C_1 = \mathsf{Enc}_{pk_R}(M_1)$ are computationally indistinguishable. Replacing $C_0$ with $C_1$ changes $\mathcal{A}$'s advantage by at most $\mathsf{negl}_2(\lambda)$.
  \item Since the ZK proof, the ciphertext, and their common commitment $c$ (together with the metadata $H(\pi)$) are the \textit{only} public outputs of the protocol visible to $\mathcal{A}$ (the audit tag is encrypted, and on-chain contract state reveals only proof verification results), the total distinguishing advantage is bounded by $\mathsf{negl}(\lambda) = \mathsf{negl}_1(\lambda) + \mathsf{negl}_2(\lambda)$. 
\end{enumerate}

\subsection{Controlled Auditability}

\begin{description}
\item[Theorem 2 (Controlled Auditability).] Under the $(t,n)$-threshold decryption scheme based on Shamir's Secret Sharing (instantiated via the Pedersen DKG and the in-the-exponent decryption of \autoref{sec:dkg}):
  \begin{enumerate}
    \item \textbf{Completeness}: Any coalition of $\ge t$ honest regulators can recover $M_{min}(\text{tx})$ from $C = \mathsf{Enc}_{pk_R}(M_{min}(\text{tx}))$.
    \item \textbf{Soundness}: Any coalition of $< t$ regulators (even if all are corrupted by $\mathcal{A}$) obtains negligible information about $M_{min}(\text{tx})$.
  \end{enumerate}
\end{description}

\noindent\textit{Proof Sketch.}
\begin{enumerate}
  \item \textbf{Completeness}: Given $t$ valid partial decryption shares $d_i = \mathsf{DecShare}(sk_i, C)$, the $\mathsf{Combine}$ algorithm applied to $\{(i, d_i)\}_{i=1}^t$ recovers the plaintext $M$. Correctness follows directly from the completeness property of the underlying $(t,n)$-threshold decryption scheme.
  \item \textbf{Soundness}: Given $t-1$ partial decryption shares $\{(i, d_i)\}_{i=1}^{t-1}$ and the ciphertext $C = \mathsf{Enc}_{pk_R}(M)$, the adversary faces two barriers: (i) the semantic security of ECIES (DDH assumption) prevents extracting $M$ from $C$ alone, and (ii) the $(t,n)$-threshold scheme ensures that fewer than $t$ shares provide negligible information about $M$. The adversary's overall advantage is therefore bounded by $\mathsf{negl}(\lambda) = \varepsilon_{\text{DDH}} + \varepsilon_{\text{threshold}}$, both negligible in the security parameter $\lambda$.
\end{enumerate}

In addition, the protocol does not automatically grant decryption capability upon satisfaction of the threshold, and legal authorization is not a cryptographic condition. Explicit legal triggers gate decryption:
\begin{itemize}
  \item \textbf{Regulatory Investigation}: Requires STR submission and internal approval from the FIU.
  \item \textbf{Law Enforcement}: Requires a judicial warrant with on-chain hash verification.
  \item \textbf{Cross-Border Request}: Requires MLAT/MMoU documentation verified by domestic judicial authorities.
\end{itemize}
These conditions are enforced through off-chain institutional procedures governing the release and combination of decryption shares, and every decryption event is recorded in the on-chain audit log. The cryptographic guarantee of Theorem 2 is therefore the threshold property itself, while compliance with legal authorization is an institutional, audit-logged control rather than a cryptographic one.

\subsection{Collusion Resistance}

\begin{description}
\item[Theorem 3 (Collusion Resistance).] Let $\mathcal{C} \subset \{R_1, \ldots, R_n\}$ be a coalition of corrupted regulators with $|\mathcal{C}| \le t-1$. Even if all members of $\mathcal{C}$ pool their key fragments $sk_i$ and all publicly available on-chain data, their advantage in recovering $M_{min}(\text{tx})$ is negligible in the security parameter $\lambda$.
\end{description}

\noindent\textit{Proof Sketch.}
By Theorem 2 (Soundness), any adversary controlling fewer than $t$ shares obtains negligible information about $M_{min}(\text{tx})$. The coalition $\mathcal{C}$ possesses at most $t-1$ shares, hence its advantage in recovering the audit plaintext is bounded by $\mathsf{negl}(\lambda)$. 

The system only becomes vulnerable when $t$ or more regulators collude - a scenario requiring coordinated action across multiple independent regulatory authorities (FIU, FSA, ISRO, judicial bodies, and/or foreign agencies), which is intentionally difficult to achieve without legal oversight and would itself constitute a prosecutable offense in most jurisdictions.

\subsection{Replay Attack Prevention}
Each audit tag includes a unique transaction hash $\mathsf{txHash}_s$ and the encrypted ciphertext $C$. The decryption process is bound to a specific transaction identifier. Replaying a previously decrypted set of partial shares against a different transaction fails because the decrypted plaintext would not match the target transaction's $\mathsf{txHash}_s$. The on-chain audit log records all decryption operations, enabling the detection of repeated decryption attempts.

\subsection{Comparative Analysis}
\begin{table}
  \caption{Comparison between Solutions}
  \label{tab:compare}
  \begin{tabular}{|l|c|c|c|c|c|}
  \toprule
  \textbf{Property} & \textbf{Transparent Bridges} & \textbf{IBC} & \textbf{Axelar} & \textbf{Privacy Coins} & \textbf{Our Framework} \\ \midrule \midrule
  Transaction Privacy     & $\times$ & $\times$ & $\times$ & $\checkmark$ & $\checkmark$ \\ \hline
  Regulatory Auditability & $\checkmark$ & $\times$ & $\times$ & $\times$ & $\checkmark$ \\ \hline
  Trust-Minimized Verification & $\times$ & $\checkmark$ & $\times$ & N/A & $\checkmark$ \\ \hline
  ZK-Based Compliance & $\times$ & $\times$ & $\times$ & $\times$ & $\checkmark$ \\ \hline
  Collusion Resistance    & N/A & N/A & N/A & N/A & $(t,n)$ threshold \\ \hline
  Decryption Accountability & always & N/A & N/A & $\times$ & conditional \\ 
  \bottomrule
  \end{tabular}
\end{table}
\autoref{tab:compare} compares our framework with existing solutions across key security and compliance properties. Transparent bridges, IBC, and Axelar each excel in different dimensions, cross-chain message passing (IBC), generalized relaying (Axelar), or audit transparency, but none provide built-in transaction privacy. Privacy coins achieve strong confidentiality but forego regulatory access. Our framework is, to the best of our knowledge, the only design that simultaneously provides transaction privacy, conditional regulatory auditability, and trust-minimized cross-chain verification.

\section{Experimental Evaluation}
This section presents a comprehensive experimental evaluation of our proposed framework. We evaluate three core components: a) the zero-knowledge proof system, b) the cross-chain verification mechanism, and c) the threshold decryption module. All experiments are conducted in realistic blockchain environments with detailed performance measurements.

\subsection{Implementation and Reproducibility}
The complete implementation, the Circom compliance circuits (commitment binding, EdDSA-Poseidon signature, AML range check, and the sorted-Poseidon Merkle blacklist), the Groth16 and PLONK setup and proving scripts, the Solidity audit/light-client contracts with a Sepolia gas-measurement script (\texttt{hardhat/}), and the Flask multi-node threshold-decryption simulator (\texttt{threshold/flask\_simulator/}), is open-sourced and publicly available at \url{https://github.com/huihengli/Balancing-Privacy-and-Compliance-in-DeFi}. The repository documents the runtime environment (Circom, snarkjs, Node.js, Python, and Solidity versions), the dependency list, and the exact commands needed to reproduce every measurement reported in this section: the cross-chain gas figures of \autoref{tab:xchainver} via \texttt{hardhat/scripts/measure\_sepolia.ts}, the decryption latencies of \autoref{tab:thresdecrypt} via \texttt{threshold/flask\_simulator/benchmark.py}, and the ablations of \autoref{tab:abl_circuit}, \autoref{tab:zksys}, and \autoref{tab:abl_component} via \texttt{build.ps1}, \texttt{benchmark\_plonk.ps1}, and the Hardhat/threshold ablation scripts. The deterministic test inputs are included, so that all reported figures can be independently verified.

\subsection{Experimental Setup}
\textbf{Hardware Environment}: All experiments were conducted on a machine equipped with an Intel Ultra 5 125H CPU (14 cores, 1.20 GHz) and 32 GB RAM, running Windows 11 Pro.\\
\textbf{Software Stacks}: The zero-knowledge proof system was implemented using Circom 2.2.3 and snarkjs 0.7.6, with Groth16 as the underlying proof system. Smart contracts were developed in Solidity and deployed on the Ethereum Sepolia testnet using Hardhat 3.3.0. The threshold decryption module was implemented in Python 3.11.9 using the Shamir library and Flask to simulate the regulator nodes.\\
\textbf{Test Data}: We generated synthetic transactions of varying levels of complexity. The compliance predicates evaluated include a) amount non-negativity, b) signature validity, c) AML threshold checking, and d) VASP blacklist non-membership, together with the commitment binding of Definition 4.

\subsection{Zero-Knowledge Proof Performance}
\label{sec:zkperf}
Our work evaluated the ZK proof system across three circuits of increasing compliance coverage. The simple circuit (8,757 constraints) verifies the commitment binding, the amount range, and an EdDSA-Poseidon signature. The medium circuit (8,825 constraints) additionally checks the AML reporting threshold. The complex circuit (19,748 constraints) incorporates VASP blacklist verification through a sorted-Poseidon Merkle non-membership proof.
\begin{table}
  \caption{Zero-Knowledge Proof Performance Metrics (30 trials)}
  \label{tab:zkproof}
  \begin{tabular}{|l|l|l|l|}
  \toprule
  \textbf{Metric} &
  \textbf{Simple Circuit} &
  \textbf{Medium Circuit} &
  \textbf{Complex Circuit} \\
  \midrule
  \midrule
  Number of Constraints        & 8,757                   & 8,825                   & 19,748                   \\ \hline
  Witness Generation Time (ms) & 189.02                  & 147.49                  & 195.96                   \\ \hline
  Proof Generation Time (ms)   & 1,060.72                & 1,054.18                & 1,412.82                 \\ \hline
  Local Verification Time (ms) & 536.63                  & 520.01                  & 479.91                   \\ \hline
  Proof Size (bytes)           & 801                     & 805                     & 803                      \\ 
  \bottomrule
  \end{tabular}
\end{table}
Key observations follow. a) Proof generation is dominated by the signature circuit: the simple and medium circuits, which differ only by an AML range check, both contain $\approx$8.8 k constraints and generate proofs in $\approx$1.05-1.06 s, whereas the complex circuit's blacklist non-membership proof (two Merkle paths) raises the count to $\approx$19.7 k and the proof time to $\approx$1.41 s. b) Witness generation ($\approx$150-196 ms) is not a bottleneck. c) On-chain verification cost is essentially independent of the number of constraints: Groth16 verification is a fixed set of pairings, so the source-chain verifier stays within the 214-229 k gas range across all tiers; local verification takes $\approx$480-537 ms. d) Proof size stays constant at $\approx$800 bytes, optimal for on-chain storage.

\noindent\textbf{Compliance-module ablation.}
\autoref{tab:abl_circuit} decomposes the three circuits as a per-module ablation of the compliance stack: each tier adds exactly one compliance predicate on top of the previous one, so the marginal cost of every module can be read off directly.
\begin{table}
  \caption{Per-module ablation of the compliance circuit. $\Delta$ denotes the marginal cost of the module added on top of the previous tier. The AML threshold is a single comparison (68 constraints) whose proof-time effect falls within measurement noise; the blacklist non-membership proof (two Poseidon Merkle paths) dominates the remaining growth. On-chain verification gas is omitted because Groth16 verification is a fixed set of pairings (214-229 k across tiers).}
  \label{tab:abl_circuit}
  \begin{tabularx}{\textwidth}{|X|r|r|r|r|}
  \toprule
  \textbf{Compliance module (cumulative)} & \textbf{Constraints} & $\Delta$ \textbf{Constraints} & \textbf{Proof time (ms)} & $\Delta$ \textbf{Proof time (ms)} \\
  \midrule
  \midrule
  Core: commitment binding + amount range + EdDSA signature & 8,757 & --- & 1,060.72 & --- \\ \hline
  + AML threshold ($\mathsf{LessThan}(64)$) & 8,825 & +68 & 1,054.18 & $\approx$0 (noise) \\ \hline
  + VASP blacklist (sorted-Poseidon Merkle non-membership) & 19,748 & +10,923 & 1,412.82 & +358.64 \\
  \bottomrule
  \end{tabularx}
\end{table}
This decomposition shows that each compliance predicate carries a quantifiable marginal cost, incurred entirely off-chain: the AML threshold adds 68 constraints with an imperceptible proof-time effect, while the blacklist non-membership proof adds 10,923 constraints and $+$358.64 ms of proof time. Because Groth16 on-chain verification performs a fixed set of pairings, none of these modules materially increases on-chain verification gas, which remains within the 214-229 k range in all tiers.

\noindent\textbf{Proof-system comparison (Groth16 vs.\ PLONK).}
To validate the choice of Groth16, we benchmarked the PLONK prover of the same snarkjs toolchain on the identical circuits and witness files. \autoref{tab:zksys} reports the comparison: Groth16 figures are 30-trial averages, and PLONK figures are steady-state averages over six runs (two batches of three); the first proof generated after PLONK setup carries a one-time proving-key preprocessing step and is excluded from the averages.
\begin{table}
  \caption{Proof-system comparison on the same circuits (snarkjs 0.7.6 reference implementations).}
  \label{tab:zksys}
  \begin{tabular}{|l|c|c|c|}
  \toprule
  \textbf{Metric} & \textbf{Groth16} & \textbf{PLONK} & \textbf{Ratio} \\
  \midrule
  \midrule
  Prove time, simple circuit (s) & 1.06 & 15.26 & $\approx$14$\times$ \\ \hline
  Prove time, medium circuit (s) & 1.05 & 13.59 & $\approx$13$\times$ \\ \hline
  Prove time, complex circuit (s) & 1.41 & 24.39 & $\approx$17$\times$ \\ \hline
  Local verification, simple circuit (ms) & 536.63 & 501.25 & $\approx$0.93$\times$ \\ \hline
  Serialized proof size, simple circuit (bytes) & 801 & 2{,}254 & $\approx$2.81$\times$ \\
  \bottomrule
  \end{tabular}
\end{table}
The comparison favors Groth16 for this workload: the reference PLONK prover is $\approx$13-17$\times$ slower (13.2-24.7 s versus 1.05-1.41 s per proof), its serialized proof is $\approx$2.8$\times$ larger, and local verification is comparable (slightly faster for PLONK). For a per-transaction compliance proof, Groth16's 1.05-1.41 s proving, $\approx$800-byte proofs, and constant On-chain verification cost ($\approx$215 k gas) remains the appropriate engineering choice; we note that optimized PLONK implementations (e.g., Plonky2) close much of this gap and leave them for future work.

\subsection{Cross-Chain Verification Performance}
We evaluated the light-client-based cross-chain verification mechanism on the Ethereum Sepolia testnet. The experiment measured three components: audit tag storage on the source chain, cross-chain proof submission, and Merkle proof verification on the target chain.
\begin{table}
  \caption{Cross-Chain Verification Performance. Gas costs exclude the Groth16 proof verification of \autoref{tab:zkproof}. The delay rows differ in measurement scope: the total cross-chain delay is the protocol-added latency measured end-to-end and excludes block confirmation time, whereas the header-synchronization and relayer-transmission durations are standalone component measurements from separate runs, include block-production waiting and network effects, and are therefore not additive.}
  \label{tab:xchainver}
  \begin{tabular}{|l|l|l|}
  \toprule
  \textbf{Component}  & \textbf{Gas Cost} & \textbf{Time (ms)} \\
  \midrule
  \midrule
  Audit Tag Storage (Source Chain) & 46,374 & ---   \\ \hline
  Light Client Verification (Target Chain) & 24,037 & --- \\ \hline
  \textbf{Total Cross-Chain Gas} & \textbf{70,411} & --- \\ \hline
  Block Header Synchronization (relayer-side, not in the total above) & 22,790 & --- \\ \hline
  \textbf{Total Cross-Chain Delay} (protocol-added, excl.\ block confirmation) & --- & \textbf{171.30} \\ \hline
  Block Header Synchronization, standalone run (incl.\ block-production waiting) & --- & 184 \\ \hline
  Relayer Transmission, standalone run (incl.\ network queueing) & --- & 333 \\
  \bottomrule
  \end{tabular}
\end{table}
In \autoref{tab:xchainver}, we observe the following. a) The cross-chain verification component (audit-tag storage and light-client verification) consumes approximately 70,411 Gas, of the same order of magnitude as a plain token transfer (our trust-the-relayer release baseline measures 46,588 gas for the bare release itself, \autoref{tab:abl_component}), and separate from the 214-229k Gas of on-chain Groth16 verification; the relayer-side block-header synchronization adds a further 22,790 Gas outside this subtotal. b) The protocol-added cross-chain delay is 171.30 ms, measured end-to-end over the verification pipeline and excluding block confirmation time (reported separately in \autoref{tab:blockconfirm}). The 184 ms header-synchronization and 333 ms relayer-transmission figures are standalone component measurements from separate runs: they include block-production waiting and network queueing that are overlapped or amortized in the end-to-end run, so the three time figures have different measurement scopes and statistical treatments and are not additive; the end-to-end total (171.30 ms) is therefore not comparable to, nor bounded by, the standalone component durations. c) Merkle proof verification using the OpenZeppelin library consumes 26,050 Gas, while our simplified light-client verification consumes 24,037 Gas, a 7.7\% optimization.
\begin{table}
  \caption{Block Confirmation Latency}
  \label{tab:blockconfirm}
  \begin{tabular}{|l|l|}
  \toprule
  \textbf{Confirmation Depth} & \textbf{Actual Time (ms)} \\
  \midrule
  \midrule
  6 Blocks                    & 5,081                     \\ \hline
  12 Blocks                   & 11,144                   \\ 
  \bottomrule
  \end{tabular}
\end{table}
As shown in \autoref{tab:blockconfirm}, block confirmation time dominates the overall transaction latency. This aligns with baseline blockchain performance and is independent of our framework's additional overhead.

\subsection{Threshold Decryption Performance}
\label{sec:thresperf}
The work also evaluated the threshold decryption mechanism with $n=5$ regulators and $t=3$ as the decryption threshold. The experiment measured the key distribution overhead and the collaborative decryption latency across 30 independent trials.
\begin{table}
  \caption{Threshold Decryption Performance (30 Trials)}
  \label{tab:thresdecrypt}
  \begin{tabular}{|l|l|}
  \toprule
  \textbf{Metric}                & \textbf{Value}        \\
  \midrule
  \midrule
  Key Distribution Time          & 547.36 ms             \\ \hline
  Average Fragment Exchange Time & 188.29 $\pm$ 20.04 ms \\ \hline
  Average Full Decryption Time   & 190.31 $\pm$ 19.53 ms     \\ \hline
  Decryption Success Rate        & 100\% (30/30)        \\
  \bottomrule
  \end{tabular}
\end{table}
Through this experiment's result shown in \autoref{tab:thresdecrypt}, indicating that: a) Key distribution is a one-time initialization cost of approximately 547 ms, which is acceptable for system deployment. b) Collaborative decryption completes in under 200 ms on average, with fragment exchange dominating the latency. c) Decryption success rate is 100\% when the threshold $t$ is satisfied, validating the correctness of the Shamir reconstruction. d) The standard deviation ($\approx$20 ms) reflects network variability, which is typical for distributed systems.

\noindent\textbf{On-chain cost of the threshold audit (analytical).}
Decryption is deliberately performed off-chain: regulators compute partial decryption shares inside HSMs and combine them over TLS, so the decryption operation itself consumes no on-chain gas by design. The only on-chain cost of an audit is the immutable audit-log entry recording each decryption event (transaction hash, timestamp, participating regulators, and the authorization hash), estimated at $\approx$25-30 k gas per logged event. For completeness, we also bound a hypothetical on-chain variant in which the contract itself combines $t$ partial shares: with the EIP-1108 bn254 precompiles, combining $t=3$ shares costs $3\times6{,}000=18$ k gas for the elliptic-curve multiplications plus a few thousand gas for Lagrange combination and key derivation, while verifying each share against its holder's public key (a DLEQ pairing check at $2\times45$ k gas per share) would raise the total to $\approx$300 k gas. Even this conservative hypothetical remains within the order of magnitude of Groth16 verification ($\approx$215 k), confirming that the audit layer imposes no disproportionate on-chain burden, and our off-chain placement removes that cost entirely.

\subsection{Performance Comparison with Baseline Solutions}
We compare our framework against recent privacy-preserving auditable cross-chain schemes, supplemented by a trustless-bridge baseline. Because re-implementing these systems is outside our scope, the figures in \autoref{tab:existcompare} are reported values from the cited works and reflect each work's own measurement scope; cross-scheme comparisons are therefore indicative rather than strictly commensurate.
\begin{table}[!htbp]
  \caption{Quantitative comparison with existing cross-chain schemes. Figures are as reported in the cited works under their respective measurement scopes; \emph{n.r.} = not reported, and \emph{ZK proof} refers to off-chain proof generation unless noted otherwise.}
  \label{tab:existcompare}
  \begin{tabularx}{\textwidth}{|X|c|c|c|c|X|}
  \toprule
  \textbf{Scheme} & \textbf{Privacy} & \textbf{Audit} & \textbf{ZK proof} & \textbf{On-chain cost} & \textbf{Overhead / notes} \\
  \midrule
  \midrule
  zkBridge (trustless baseline) \cite{xie_2022_zkbridge} & $\times$ & $\times$ & $\approx$20 s (Cosmos header) & $<$230 k gas (verify) & Trustless; no privacy or audit \\ \midrule
  VeilAudit \cite{Qiao2025VeilAudit} & \checkmark & $\triangle$ & 1.1-1.2 s & +695-872 k over a 116 k baseline & AOL tag + threshold reveal; $\approx$20 ms verify \\ \midrule
  privXCA \cite{Wang2026privXCA} & \checkmark & \checkmark & recursive (Plonky2) & 174 k (client) / 124 k (audit state) & $\approx$3 s window audit ($\approx$810$\times$ faster) \\ \midrule
  Mystiko \cite{Li2025An} & \checkmark & \checkmark & 5.2-5.8 s (withdrawal) & 644 k avg per transfer & Rollup batching reduces cost $\approx$90\% \\ \midrule
  CPAP \cite{Jiang2025Cross} & \checkmark & $\triangle$ & n.r. & $\approx$120 k (20-byte msg) & $\approx$4 TPS; assumes honest bridge \\ \midrule
  Notary + ZKP \cite{Wu2025Privacy} & \checkmark & $\triangle$ & 5-15 s & n.r. (local-only eval.) & Notary-only accountability \\ \midrule
  Anoma \cite{khalniyazova_2023_anoma} & \checkmark & $\times$ & n.r. & n.r. & No audit channel; no published per-transfer figures \\ \midrule
  IBC privacy extensions \cite{zamyatin_2021_sok} & $\triangle$ & $\times$ & n.r. & n.r. & Privacy bolted onto IBC; no published comparable figures \\ \midrule
  Our framework & \checkmark & \checkmark & 1.05-1.41 s & 214-229 k (ZK verify) + $\approx$70 k (cross-chain) & +171 ms latency; threshold decrypt $<$200 ms \\
  \bottomrule
  \end{tabularx}
\end{table}
As \autoref{tab:existcompare} shows, our framework's off-chain proof generation (1.05-1.41 s) is on par with VeilAudit (1.1-1.2 s) and markedly faster than Mystiko (5.2-5.8 s) and the notary scheme (5-15 s), and its on-chain cost (214-229 k gas for Groth16 verification plus $\approx$70 k gas for the cross-chain component) is competitive with Mystiko's 644 k and privXCA's per-update costs. Privacy-first designs such as Anoma and the IBC privacy extensions offer no regulator-facing audit channel and publish no comparable per-transfer figures (marked n.r. in \autoref{tab:existcompare}). Unlike zkBridge, which achieves trustless verification without privacy or audit, and unlike the aggregate-audit and notary-only designs, our framework is, to the best of our knowledge, the only scheme in this class that simultaneously provides transaction privacy, threshold-gated minimal-disclosure audit, and trust-minimized light-client verification. On the Ethereum Sepolia testnet with synthetic transactions, the measured end-to-end overhead remains modest ($\approx$171 ms of additional latency and $\approx$70 k gas, both dominated by block confirmation rather than by our protocol); extending these results to additional chains and real regulatory workflows remains future work.

\noindent\textbf{Component-level ablation.}
\autoref{tab:abl_component} isolates the marginal cost of each of the three building blocks by measuring a variant of the framework with that component removed, and records the security property lost by the removal.
\begin{table}[!htbp]
  \caption{Component-level ablation: the marginal cost of each building block and the property lost when it is removed. ZK figures are from \autoref{tab:zkproof}; the light-client delta is the measured gas difference between a Merkle-verifying release and a trust-the-relayer release (baseline 46,588 gas) of the same contract; threshold figures are loopback share-exchange measurements, with the Flask multi-node exchange of \autoref{tab:thresdecrypt} in parentheses.}
  \label{tab:abl_component}
  \begin{tabularx}{\textwidth}{|l|X|X|}
  \toprule
  \textbf{Component} & \textbf{Marginal cost (measured)} & \textbf{Property lost if removed} \\
  \midrule
  \midrule
  ZK compliance module & 214-229 k gas on-chain + 1.05-1.41 s off-chain proving & Transaction privacy and publicly verifiable compliance \\ \midrule
  Light-client verification & 7,826 gas marginal over the 46,588 gas trust-the-relayer release baseline & Trust-minimized verification (must trust the relayer) \\ \midrule
  Threshold decryption & 0.01 ms computation + 8.47 ms loopback share exchange (188.29 ms with the Flask multi-node setup) & Collusion resistance (any single key holder could decrypt) \\
  \bottomrule
  \end{tabularx}
\end{table}
This decomposition quantifies the price of each guarantee: privacy-preserving compliance costs 214-229 k gas and roughly one second of off-chain proving; trust-minimized verification costs 7,826 gas against trusting a relayer; and threshold control costs the share-exchange latency (8.47 ms over loopback, 188.29 ms in the multi-node Flask setup) while the cryptographic reconstruction itself is negligible (0.01 ms). Each cost is therefore small, measurable, and confined to the layer that provides the corresponding property.

\clearpage
\section{Discussion and Future Work}
\subsection{Regulatory Compatibility}
Our framework aligns with key regulatory requirements. The encrypted audit tag contains only the minimum information required by the FATF Travel Rule and MiCA, ensuring compliance without unnecessary data exposure. The threshold-based audit mechanism respects legal due process, since decryption requires both a judicial warrant (for law enforcement scenarios) and collaboration from at least $t$ regulators. This dual control prevents unilateral surveillance while enabling lawful oversight. However, real-world deployment faces policy challenges. Cross-border regulatory coordination requires harmonized legal frameworks for key fragment management. We envision that international bodies such as the Financial Action Task Force could establish standards for key regulatory ceremonies, similar to those in existing mutual legal assistance treaties.
\subsection{Limitations and Future Directions}
While our ZK circuits achieve proof generation in $\approx$1.0-1.4 s, further optimization is possible. The current circuits support basic compliance checks (amount range, signature, blacklist). More complex predicates, such as historical transaction analysis, would increase constraint count and proof time. Our prototype targets EVM-compatible chains, as stated in the assumptions of \autoref{sec:assumptions}. The chain-independent components, the Groth16 circuits, the encrypted audit tag format, the threshold decryption protocol, and the audit contract logic, carry over unchanged. In contrast, the light-client contract and the Merkle-tree and header-verification conventions must be re-implemented per chain. Extending to non-EVM chains (e.g., Bitcoin via SPV-style header verification, or Solana) therefore requires chain-specific light-client implementations of varying complexity, which we leave as future work. Our framework currently assumes an honest majority during DKG execution (i.e., fewer than $t$ malicious regulators during key generation). While this is strictly weaker than trusting a single KGC, future work could explore publicly verifiable DKG protocols that provide security even under a dishonest majority, at the cost of additional communication rounds. The minimum audit information set, while privacy-preserving, may still reveal more than necessary for certain investigations. Future work could explore differential privacy techniques to add controlled noise to audit data, providing an additional layer of privacy protection for routine compliance monitoring. For blacklist liveness, the Merkle root approach requires that the blacklist authority (e.g., FIU) update the on-chain root promptly. A malicious authority could delay updates, allowing blacklisted VASPs to continue transacting. This is a governance, not cryptographic, limitation. In practice, the authority's update schedule could be enforced via smart contract timelocks or multi-signature thresholds. Future work could explore decentralized blacklist management using on-chain governance or oracle-based attestations. Finally, the threshold-decryption evaluation is simulator-based: the decryption latency of \autoref{tab:thresdecrypt} was measured with a Python/Flask multi-node simulator rather than a production deployment, and because decryption is intentionally off-chain it contributes no gas. The evaluation therefore does not capture production-grade regulator infrastructure (HSM latency, real-world network topology, and key-ceremony logistics); an analytical on-chain cost model is provided in \autoref{sec:thresperf}, and a production deployment remains future work.

\section{Conclusion}
This work presented a privacy-preserving and auditable cross-chain transaction framework that reconciles the fundamental tension between user privacy and regulatory compliance. By integrating zero-knowledge proofs, threshold cryptography, and light-client verification, our design enables transaction details to remain confidential during normal operation, with only ZK proofs publicly verifiable. The regulatory bodies' access to audit information requires both legal authorization and collaborative decryption by at least $t$ regulators. Asset transfers are validated via cryptographic proofs without reliance on centralized bridges.\par
Experimental evaluation on the Ethereum Sepolia testnet with synthetic transactions demonstrates the framework's practicality. ZK proof generation completes in $\approx$1.0-1.4 s; cross-chain verification adds approximately 170 ms of latency and 70k Gas (audit-tag storage plus light-client verification), on top of 214-229k Gas for on-chain Groth16 verification; and threshold decryption finishes in under 200 ms. These figures indicate feasibility for compliance-sensitive DeFi applications, where block confirmation times dominate overall latency, while deployment on additional chains and with real regulatory workflows remains future work.

\section*{Acknowledgement}
The authors would like to express their sincere gratitude to the following professors for their valuable contributions to this work. We are especially grateful to Prof. Mingzhe Li (Great Bay University) for his detailed and professional suggestions during the revision process, which significantly improved the quality of this paper. We also thank Prof. Jing Liu (Guangdong University of Finance and Economics) for reviewing the manuscript and providing insightful feedback. We further extend our thanks to all colleagues and friends who supported this research.

\section*{Declaration of generative AI and AI-assisted technologies in the manuscript preparation process}
During the preparation of this work, the authors used ChatGPT for clarity improvement. The authors reviewed and edited the output as needed and take full responsibility for the content of the published article. 

\printcredits

\bibliographystyle{model1-num-names}

\bibliography{2026pnr_r1}

\begin{thebibliography}{38}
\expandafter\ifx\csname natexlab\endcsname\relax\def\natexlab#1{#1}\fi
\providecommand{\bibinfo}[2]{#2}
\ifx\xfnm\relax \def\xfnm[#1]{\unskip,\space#1}\fi
\bibitem[{Lee(2021)}]{lee_2021_crypto}
\bibinfo{author}{I.~Lee}, \bibinfo{title}{Crypto market cap doubles to \$2 trillion as rally surges}, \bibinfo{year}{2021}.
\bibitem[{Deng et~al.(2025)Deng, Wang, Wang, Wang, Zhu, and Zhang}]{deng_2025_a}
\bibinfo{author}{H.~Deng}, \bibinfo{author}{Z.~Wang}, \bibinfo{author}{Y.~Wang}, \bibinfo{author}{L.~Wang}, \bibinfo{author}{L.~Zhu}, \bibinfo{author}{C.~Zhang},
\newblock \bibinfo{title}{A secure cross-account audit scheme for cross-chain transactions},
\newblock in: \bibinfo{booktitle}{International Conference on Algorithms and Architectures for Parallel Processing}, volume \bibinfo{volume}{15253}, \bibinfo{organization}{Lecture Notes in Computer Science}, \bibinfo{publisher}{Springer Nature Singapore}, \bibinfo{year}{2025}, pp. \bibinfo{pages}{334--350}.
\bibitem[{Feng et~al.(2025)Feng, Wu, Hu, and Yao}]{feng_2025_secure}
\bibinfo{author}{L.~Feng}, \bibinfo{author}{P.~Wu}, \bibinfo{author}{K.~Hu}, \bibinfo{author}{S.~Yao},
\newblock \bibinfo{title}{Secure cross-chain interactions: a relay chain framework with privacy and auditability solutions},
\newblock \bibinfo{journal}{Cluster Computing} \bibinfo{volume}{28} (\bibinfo{year}{2025}).
\bibitem[{Chen et~al.(2024)Chen, Lu, and Chen}]{chen_2024_crosschain}
\bibinfo{author}{Z.~Chen}, \bibinfo{author}{Z.~Lu}, \bibinfo{author}{J.~Chen},
\newblock \bibinfo{title}{Cross-chain trusted information match scheme with privacy-preserving and auditability},
\newblock in: \bibinfo{booktitle}{Blockchain Technology and Emerging Applications}, volume \bibinfo{volume}{577}, \bibinfo{organization}{Lecture Notes of the Institute for Computer Sciences, Social Informatics and Telecommunications Engineering}, \bibinfo{publisher}{Springer Nature Switzerland}, \bibinfo{year}{2024}, pp. \bibinfo{pages}{94--114}.
\bibitem[{Li et~al.(2024)Li, Qi, Xu, Zhu, Zhou, Wen, and Xiang}]{li_2024_blockchain}
\bibinfo{author}{N.~Li}, \bibinfo{author}{M.~Qi}, \bibinfo{author}{Z.~Xu}, \bibinfo{author}{X.~Zhu}, \bibinfo{author}{W.~Zhou}, \bibinfo{author}{S.~Wen}, \bibinfo{author}{Y.~Xiang},
\newblock \bibinfo{title}{Blockchain cross-chain bridge security: Challenges, solutions, and future outlook},
\newblock \bibinfo{journal}{Distributed Ledger Technologies: Research and Practice} \bibinfo{volume}{4} (\bibinfo{year}{2024}).
\bibitem[{Xie et~al.(2022)Xie, Zhang, Cheng, Zhang, Zhang, Jia, Boneh, and Song}]{xie_2022_zkbridge}
\bibinfo{author}{T.~Xie}, \bibinfo{author}{J.~Zhang}, \bibinfo{author}{Z.~Cheng}, \bibinfo{author}{F.~Zhang}, \bibinfo{author}{Y.~Zhang}, \bibinfo{author}{Y.~Jia}, \bibinfo{author}{D.~Boneh}, \bibinfo{author}{D.~Song},
\newblock \bibinfo{title}{zkbridge: Trustless cross-chain bridges made practical},
\newblock in: \bibinfo{booktitle}{Proceedings of the 2022 ACM SIGSAC Conference on Computer and Communications Security}, \bibinfo{publisher}{ACM}, \bibinfo{year}{2022}, pp. \bibinfo{pages}{3003--3017}.
\bibitem[{Nikolaeva(2022)}]{nikolaeva_2022_axelar}
\bibinfo{author}{O.~Nikolaeva}, \bibinfo{title}{Axelar network: Connecting applications with blockchain ecosystems}, \bibinfo{year}{2022}.
\bibitem[{Liu et~al.(2024)Liu, Luo, Yan, Izhikevich, Grant, Stefan, Voelker, and Savage}]{liu_2024_accounting}
\bibinfo{author}{E.~Liu}, \bibinfo{author}{E.~Luo}, \bibinfo{author}{J.~C. Yan}, \bibinfo{author}{K.~Izhikevich}, \bibinfo{author}{S.~Grant}, \bibinfo{author}{D.~Stefan}, \bibinfo{author}{G.~M. Voelker}, \bibinfo{author}{S.~Savage},
\newblock \bibinfo{title}{Count of monte crypto: Accounting-based defenses for cross-chain bridges},
\newblock \bibinfo{journal}{arXiv preprint arXiv:2410.01107}  (\bibinfo{year}{2024}).
\bibitem[{Belchior et~al.(2024)Belchior, Torres, Pfannschmidt, Vasconcelos, and Correia}]{belchior_2024_bungee}
\bibinfo{author}{R.~Belchior}, \bibinfo{author}{L.~Torres}, \bibinfo{author}{J.~Pfannschmidt}, \bibinfo{author}{A.~Vasconcelos}, \bibinfo{author}{M.~Correia},
\newblock \bibinfo{title}{Bungee: Dependable blockchain views for interoperability},
\newblock \bibinfo{journal}{Distributed Ledger Technologies: Research and Practice} \bibinfo{volume}{3} (\bibinfo{year}{2024}).
\bibitem[{Zamyatin et~al.(2021)Zamyatin, Al-Bassam, Zindros, Kokoris-Kogias, Moreno-Sanchez, Kiayias, and Knottenbelt}]{zamyatin_2021_sok}
\bibinfo{author}{A.~Zamyatin}, \bibinfo{author}{M.~Al-Bassam}, \bibinfo{author}{D.~Zindros}, \bibinfo{author}{E.~Kokoris-Kogias}, \bibinfo{author}{P.~Moreno-Sanchez}, \bibinfo{author}{A.~Kiayias}, \bibinfo{author}{W.~J. Knottenbelt},
\newblock \bibinfo{title}{Sok: Communication across distributed ledgers},
\newblock \bibinfo{journal}{Financial Cryptography and Data Security}  (\bibinfo{year}{2021}) \bibinfo{pages}{3--36}.
\bibitem[{Khalniyazova(2023)}]{khalniyazova_2023_anoma}
\bibinfo{author}{Y.~Khalniyazova}, \bibinfo{title}{Privacy in tents: An overview of private solving strategies}, \bibinfo{year}{2023}.
\bibitem[{Sasson et~al.(2014)Sasson, Chiesa, Garman, Green, Miers, Tromer, and Virza}]{sasson_2014_zerocash}
\bibinfo{author}{E.~B. Sasson}, \bibinfo{author}{A.~Chiesa}, \bibinfo{author}{C.~Garman}, \bibinfo{author}{M.~Green}, \bibinfo{author}{I.~Miers}, \bibinfo{author}{E.~Tromer}, \bibinfo{author}{M.~Virza},
\newblock \bibinfo{title}{Zerocash: Decentralized anonymous payments from bitcoin},
\newblock in: \bibinfo{booktitle}{2014 IEEE Symposium on Security and Privacy}, \bibinfo{publisher}{{IEEE} Computer Society}, \bibinfo{year}{2014}, pp. \bibinfo{pages}{459--474}.
\bibitem[{J and S(2026)}]{J2026Monero}
\bibinfo{author}{C.~J}, \bibinfo{author}{D.~S},
\newblock \bibinfo{title}{Monero privacy mining architecture and regulatory compliance and a systematic review},
\newblock \bibinfo{journal}{Discover Computing} \bibinfo{volume}{29} (\bibinfo{year}{2026}).
\bibitem[{{AML Network}(2025)}]{amlnetwork_2025_kyc}
\bibinfo{author}{{AML Network}}, \bibinfo{title}{What is know your customer in anti-money laundering? - aml network}, \bibinfo{year}{2025}.
\bibitem[{{FATF}(2021)}]{fatf_2021_updated}
\bibinfo{author}{{FATF}}, \bibinfo{title}{Updated guidance for a risk-based approach to virtual assets and virtual asset service providers}, \bibinfo{year}{2021}.
\bibitem[{{Chainalysis}(2024)}]{chainalysis_2024_blockchain}
\bibinfo{author}{{Chainalysis}}, \bibinfo{title}{Blockchain intelligence - chainalysis}, \bibinfo{year}{2024}.
\bibitem[{Li et~al.(2025)Li, D'Angelo, Tang, Fang, and Gong}]{Li2025An}
\bibinfo{author}{A.~Li}, \bibinfo{author}{G.~D'Angelo}, \bibinfo{author}{S.-K. Tang}, \bibinfo{author}{F.~Fang}, \bibinfo{author}{B.~Gong},
\newblock \bibinfo{title}{An auditable confidentiality protocol for blockchain transactions},
\newblock \bibinfo{journal}{Blockchain: Research and Applications}  (\bibinfo{year}{2025}) \bibinfo{pages}{100380}.
\bibitem[{Qiao et~al.(2025)Qiao, Dong, and Gondal}]{Qiao2025VeilAudit}
\bibinfo{author}{M.~Qiao}, \bibinfo{author}{H.~Dong}, \bibinfo{author}{I.~Gondal}, \bibinfo{title}{Veilaudit: Breaking the deadlock between privacy and accountability across blockchains}, \bibinfo{year}{2025}.
\bibitem[{Wang et~al.(2026)Wang, Wu, Chen, and Han}]{Wang2026privXCA}
\bibinfo{author}{J.~Wang}, \bibinfo{author}{C.~Wu}, \bibinfo{author}{Y.~Chen}, \bibinfo{author}{W.~Han},
\newblock \bibinfo{title}{privxca: An efficient and privacy-preserving auditing architecture for cross-chain transfers},
\newblock \bibinfo{journal}{Journal of Systems Architecture} \bibinfo{volume}{171} (\bibinfo{year}{2026}) \bibinfo{pages}{103662}.
\bibitem[{Wu et~al.(2025)Wu, Zhang, Chen, and Wang}]{Wu2025Privacy}
\bibinfo{author}{X.~Wu}, \bibinfo{author}{T.~Zhang}, \bibinfo{author}{L.~Chen}, \bibinfo{author}{Z.~Wang},
\newblock \bibinfo{title}{Privacy-preserving cross-chain asset transfers using notary schemes and zero-knowledge proofs},
\newblock \bibinfo{journal}{Cluster Computing} \bibinfo{volume}{28} (\bibinfo{year}{2025}).
\bibitem[{Jiang et~al.(2025)Jiang, Liu, Liu, and Zhu}]{Jiang2025Cross}
\bibinfo{author}{P.~Jiang}, \bibinfo{author}{J.~Liu}, \bibinfo{author}{Q.~Liu}, \bibinfo{author}{L.~Zhu},
\newblock \bibinfo{title}{Cross-chain privacy-preserving auditable protocol},
\newblock \bibinfo{journal}{Proceedings of the 7th ACM International Symposium on Blockchain and Secure Critical Infrastructure}  (\bibinfo{year}{2025}) \bibinfo{pages}{1--8}.
\bibitem[{FATF(2025)}]{fatf_2025_fatf}
\bibinfo{author}{FATF}, \bibinfo{title}{Fatf updates standards on recommendation 16 on payment transparency}, \bibinfo{year}{2025}.
\bibitem[{Limited(2020)}]{ivxsuklimited_2020_6amld}
\bibinfo{author}{I.~U. Limited}, \bibinfo{title}{6amld: 6th anti money laundering directive}, \bibinfo{year}{2020}.
\bibitem[{Deane(2025)}]{deane_2025_a}
\bibinfo{author}{R.~Deane}, \bibinfo{title}{A complete guide to aml regulations around the world}, \bibinfo{year}{2025}.
\bibitem[{{Chainalysis}(2024)}]{chainalysisteam_2024_fatfs}
\bibinfo{author}{{Chainalysis}}, \bibinfo{title}{Fatf's report on recommendation 15: What it means}, \bibinfo{year}{2024}.
\bibitem[{{AML Network}(2025)}]{amlnetwork_2025_aml}
\bibinfo{author}{{AML Network}}, \bibinfo{title}{What are the global anti-money laundering laws?}, \bibinfo{year}{2025}.
\bibitem[{FATF(2025)}]{fatf_2025_asset}
\bibinfo{author}{FATF}, \bibinfo{title}{Fatf urges stronger global action to address illicit finance risks in virtual assets}, \bibinfo{year}{2025}.
\bibitem[{Schwind-Wagner(2025)}]{schwindwagner_2025_fatf}
\bibinfo{author}{B.~Schwind-Wagner}, \bibinfo{title}{Fatf | r.29 financial intelligence units}, \bibinfo{year}{2025}.
\bibitem[{FATF/OECD(2012)}]{fatfoecd_2012_international}
\bibinfo{author}{FATF/OECD}, \bibinfo{title}{International standards on combating money laundering and the financing of terrorism \& proliferation}, \bibinfo{year}{2012}.
\bibitem[{Parliament and of~the European~Union(2023)}]{europeanparliament_2023_regulation}
\bibinfo{author}{E.~Parliament}, \bibinfo{author}{C.~of~the European~Union}, \bibinfo{title}{Regulation (eu) 2023/1114 of the european parliament and of the council of 31 may 2023 on markets in crypto-assets, and amending regulations (eu) no 1093/2010 and (eu) no 1095/2010 and directives 2013/36/eu and (eu) 2019/1937}, \bibinfo{year}{2023}.
\bibitem[{Parliament and of~the European~Union(2018)}]{europeanparliament_2018_directive}
\bibinfo{author}{E.~Parliament}, \bibinfo{author}{T.~C. of~the European~Union}, \bibinfo{title}{Directive (eu) 2018/1673 of the european parliament and of the council of 23 october 2018 on combating money laundering by criminal law}, \bibinfo{year}{2018}.
\bibitem[{Commission(2023)}]{theeuropeancommission_2023_commission}
\bibinfo{author}{T.~E. Commission}, \bibinfo{title}{Commission implementing regulation (eu) 2023/1111 of 6 june 2023 amending regulation (ec) no 474/2006 as regards the list of air carriers banned from operating or subject to operational restrictions within the union}, \bibinfo{year}{2023}.
\bibitem[{Parliament and of~the European~Union(2014)}]{europeanparliament_2014_directive}
\bibinfo{author}{E.~Parliament}, \bibinfo{author}{T.~C. of~the European~Union}, \bibinfo{title}{Directive 2014/62/eu of the european parliament and of the council of 15 may 2014 on the protection of the euro and other currencies against counterfeiting by criminal law, and replacing council framework decision 2000/383/jha}, \bibinfo{year}{2014}.
\bibitem[{of~International Affairs Criminal Division U.S. Department~of Justice(2022)}]{officeofinternationalaffairscriminaldivisionusdepartmentofjustice_2022_mutual}
\bibinfo{author}{O.~of~International Affairs Criminal Division U.S. Department~of Justice}, \bibinfo{title}{Mutual legal assistance treaties of the united states}, \bibinfo{year}{2022}.
\bibitem[{of~Securities~Commissions(2016)}]{internationalorganizationofsecuritiescommissions_2016_2013}
\bibinfo{author}{I.~O. of~Securities~Commissions}, \bibinfo{title}{2013 list of non-signatories to the mmou}, \bibinfo{year}{2016}.
\bibitem[{of~The European~Communities(2000)}]{thecommissionoftheeuropeancommunities_2000_2000800ec}
\bibinfo{author}{T.~C. of~The European~Communities}, \bibinfo{title}{2000/800/ec: Commission decision of 7 december 2000 authorising the member states to permit temporarily the marketing of vine propagating material not satisfying the requirements of council directive 68/193/eec}, \bibinfo{year}{2000}.
\bibitem[{Office(2024)}]{homeoffice_2024_0052024}
\bibinfo{author}{H.~Office}, \bibinfo{title}{005/2024: Economic crime and corporate transparency act - cryptoasset forfeiture provisions chapters 3c to 3f}, \bibinfo{year}{2024}.
\bibitem[{Services and the Treasury Bureau of~the Hong Kong Special Administrative Region~Government(2018)}]{financialservicesandthetreasurybureauofthehongkongspecialadministrativeregiongovernment_2018_antimoney}
\bibinfo{author}{F.~Services}, \bibinfo{author}{the Treasury Bureau of~the Hong Kong Special Administrative Region~Government}, \bibinfo{title}{Anti-money laundering and counter-terrorist financing ordinance}, \bibinfo{year}{2018}.

\end{thebibliography}

\end{document}